\documentclass[aps,prl,floatfix,twocolumn,nofitinbib,hyperref=pdftex,citeautoscript]{revtex4}

\usepackage{epsfig}
\usepackage{svg}

\usepackage{amsmath,bm}
\usepackage{amssymb}
\usepackage{amsfonts}
\usepackage{braket}
\usepackage{dsfont}
\usepackage{comment,color}
\usepackage{lipsum}
\usepackage{hyperref}

\hypersetup{colorlinks=true,linkcolor=blue,anchorcolor=blue,citecolor=blue,filecolor=blue,urlcolor=blue,bookmarksnumbered=true,pdfview=FitB}

\newcommand{\nn}{\nonumber\\}

\begin{document}
\title{Floquet Exceptional Topological Insulator}

\author{Gaurab Kumar Dash}
\author{Subhajyoti Bid}
\author{Manisha Thakurathi}

\affiliation{Department of Physics, Indian Institute of Technology Delhi, Hauz Khas, New Delhi, India 110016}

\date{\today}

\begin{abstract}
    We propose a novel way of modulating exceptional topology by implementing Floquet engineering in non-hermitian (NH) systems. We introduce Floquet exceptional topological insulator which results from shining light on a conventional three-dimensional NH topological insulator. Light-matter interaction facilitates the quantum phases of matter to exhibit a novel phenomenon, where, the point gaps in the bulk host surface states. These distinct surface states either fill the point gap in the complex eigenspectrum or  exhibit exceptional points in the presence of a magnetic field. We also highlight the existence of a quantum anomaly generated by photo-induced modulation. The existence of the Floquet biorthogonal Chern number and spectral winding number show that the momentum slices exhibit NH skin effect, even though the system as a whole does not. We also employ wave-dynamics evolution to illustrate the NH surface skin effect.
\end{abstract}

\maketitle
\textit{Introduction.\textemdash}
The non-hermitian (NH) topological phases have been the center of attraction for theoretical as well as experimental studies \cite{RevModPhys.93.015005,Sbid}. Many experimentally realizable models of NH systems have been proposed in condensed matter physics hosting exceptional points (EPs), lines, rings, and nodal planes \cite{Soori, Duan}. Recently, NH generalization of topological insulators (TI) known as Exceptional TIs has been studied where the surface hosts either a 2D band structure with a single EP (a point where both eigenenergies and eigenvectors coalesce) or a single band, which represents a vortex \cite{denner2021exceptional}. This is obtained either by varying the imbalance between the g-factors of the orbitals or the magnetic field strength while keeping it isotropic along [111] direction.
\begin{figure}[t]
    \centering

     \includegraphics[width=0.65\linewidth]{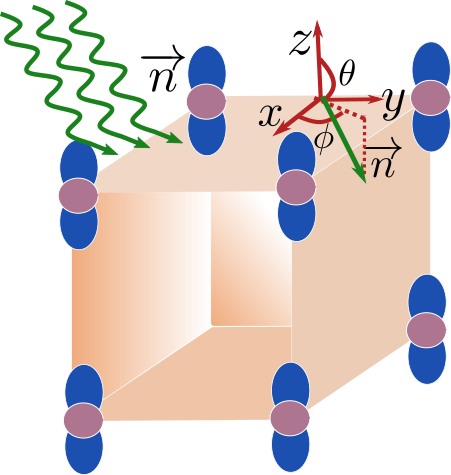}
 
    \caption{Schematic picture of a cubic lattice with $s$ and $p$ orbital at each site where a CPL light is irradiated along $\vec n$ direction with $\theta$ and $\phi$ being the angles of polarization.}
    \label{fig1}
\end{figure}
On the other hand, the Floquet engineering of the topological phases in Hermitian \cite{okatashi, Sen1, weylFlo, AS1, JK1, FloquetTop, MT1, Floquetsum, JK2, DisFlo, AS2, Sen2}, as well as NH systems, has defined exotic ramifications in stroboscopic limits which are otherwise not realizable in their static counterparts\cite{Burkov1, Burkov2}. Therefore, Floquet engineering of exceptional topology has captured the desirable attention \cite{conEP}. 

In this letter, we interlink these two ideas and introduce a 3D Floquet exceptional topological insulator (FETI). We also aim to answer the following questions. Is it possible to achieve the ETI phase without an external magnetic field, convert a hugely defective point into non-defective points, and modulate exceptional topology using Floquet theory? In order to answer these questions, we start with conventional 3D NHTI and shine circularly polarised light (CPL) on it \cite{wang2013observation}. The system hosts a central point gap in the bulk which can be transitioned into a central line gap by tuning the amplitude of the vector potential of CPL. The point gaps, which are characterized by the quantized value of topological invariant in the bulk, host an infernal point  \cite{denner2021exceptional}  leading to the defectiveness of the system in the static part. However, in the stroboscopic phase, the system experiences a photo-induced pseudo-magnetic field and hosts robust non-defective doubly degenerate surface states localized at opposite surfaces. The doubly degenerate surface states fill the point gap and exhibit different dispersion relations.  The surface state appears as a single sheet which represents a vortex without an anti-vortex partner giving rise to the quantum anomaly in 3D NHTI systems. This sheet can be shown to establish homotopy with a torus-shaped 2D Brillouin zone (BZ). 

However, when the system is subjected to a constant and isotropic magnetic field, it hosts a single second-order EP on the surface of the 3D NHTI models. In the presence of CPL, it also experiences a photo-induced  pseudo-magnetic field which couples with the external magnetic field \cite{bukov2015universal} to modulate the EP on the surface. Thus, a constant magnetic field in the static phase gains dynamics in the stroboscopic picture thereby dramatically photo-modulating the Land\'e-g-factor. The system possesses NH surface skin effect (NHSSE) \cite{NHSE,NHSEO} in both, static as well as dynamic phases. Here we explain this surface quenching phenomenon with the help of wave dynamics\cite{Dynweyl,xiao2020non} without analyzing the photovoltaic and chiral transport phenomenons associated with it.

\textit{Static model\textemdash}
We start with the quantum mechanical model of spin-full 3D NHTI, given by following Hamiltonian,
\begin{align}\label{1}
    H_0(k) =\sum_{j=x,y,z}\big[&\left(\cos k_j - M\right)\tau_z\sigma_0 + \lambda \sin k_j\tau_x\sigma_j \big]\nn
    &+ i\,\delta\,\tau_x\sigma_0.
\end{align}

 In the tight-binding model, the system has $s$ and $p$ orbital at each lattice site \cite{denner2021exceptional, Dynweyl} (implying four degrees of freedom at each site) see Fig. \ref{fig1}(a), $\sigma_\mu$ and $\tau_\mu$ denotes the Pauli matrices acting independently of spin and orbit degrees of freedom respectively where $\mu = 0,1,2,3$. Here, $M$ accounts for the band inversion for the $s$ and $p$ orbital, $\lambda$ controls the intrinsic spin-orbit coupling, $\delta$ denotes the NH contribution either from electron-phonon scattering or via short-lived $f$-electron coupling to the $s$ and $p$ orbitals \cite{denner2021exceptional}. We note that the Hermitian counterpart of the above model mentioned in Eq. \ref{1} resembles a four-band model of  3D TI hosting a Dirac node. The static Hamiltonian hosts two kinds of gaps in the bulk due to its NH property, point gap for $\lvert M-3\rvert\leq\delta$ as shown in Fig. \ref{fig1}(b) and  line gap for $\lvert M-3\rvert\geq\delta$.

\begin{figure}[t]
    \centering
   
\begin{tabular}{c c}
 \includegraphics[width=0.48\linewidth]{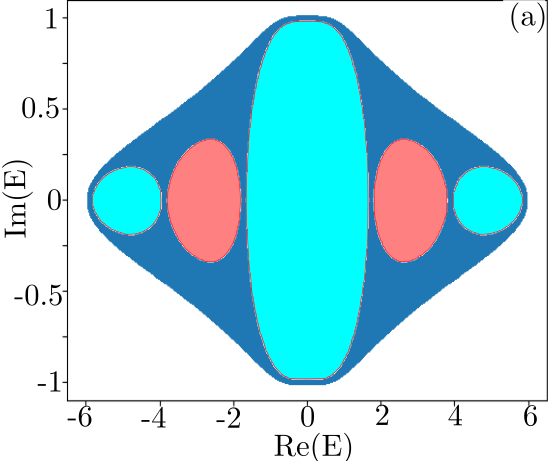}&
    \includegraphics[width=0.48\linewidth]{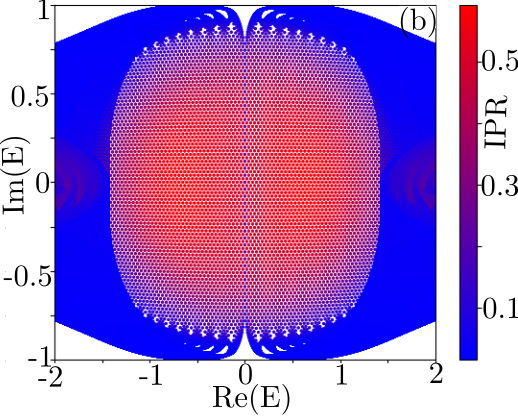}\\
     \includegraphics[width=0.55\linewidth]{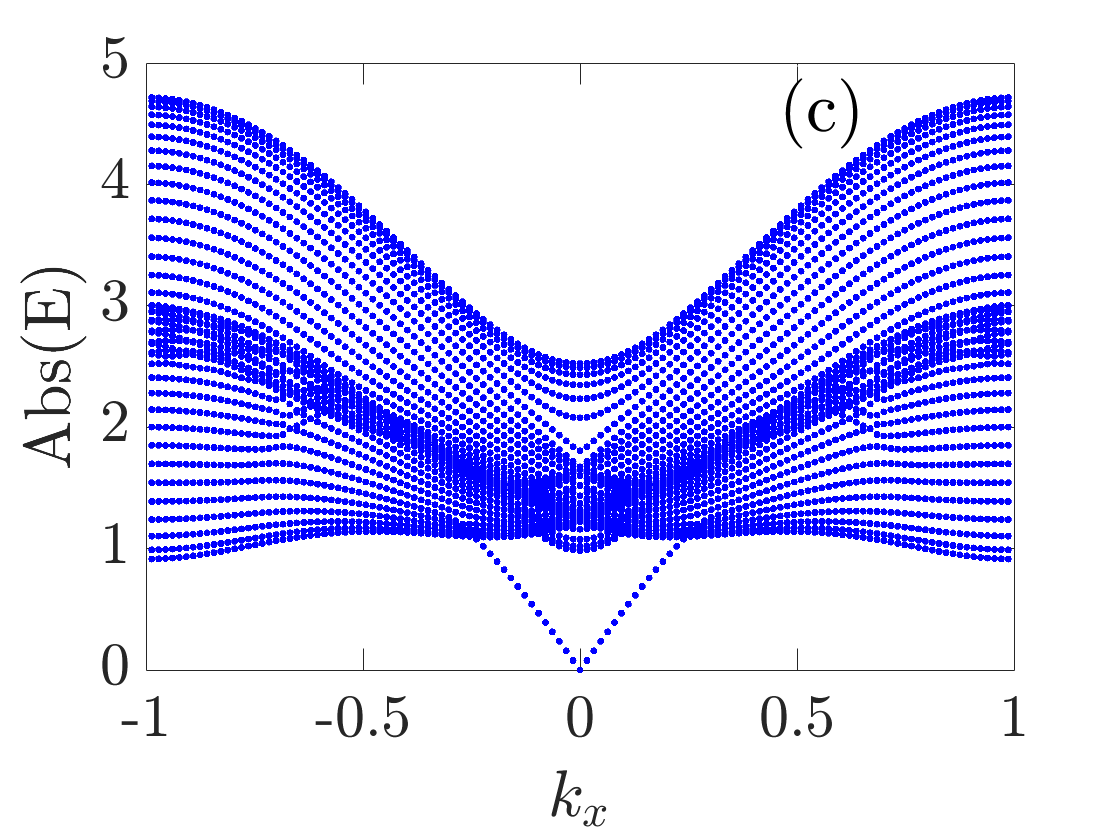}
     \hspace*{-0.4cm}
    & \includegraphics[width=0.5\linewidth]{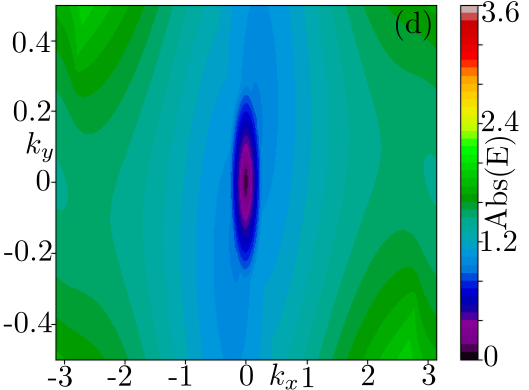}
    \end{tabular}
    \caption{(a) Bulk complex eigenspectrum  for point gap region satisfying $\lvert M+\frac{A^2}{2}-3\rvert\leq\delta$, cyan (pink) color highlights the value of $W_{3D}=1 (-2)$. (b) represents the central point gap region of complex eigenspectrum for OBC along the $z$ axis. The degenerate surface states fill the point gap completely and appear as a single sheet structure. (c) Plot of Abs(E) as a function of $k_x$ for $k_y = 0$,  hosting a surface Dirac node at $k_x = 0$. Notably, when $k_y$ is non-zero the surface is gapped and the states fill the Dirac cone completely as appears in the contour plot (d). System parameters are $M=2.5$, $\lambda = 1$, $\delta=1$, $A_0^2/\omega =1/5 $ and momentum resolution $\Delta k=\pi/400$.}
    \label{fig2}
\end{figure}

\textit{Floquet exceptional topological insulator (FETI).\textemdash}
We irradiate CPL \cite{A} along the direction $\vec{n} = (\sin\theta\cos\phi, \, \sin\theta\sin\phi, \, \cos\theta)$, where, $\theta$ and $\phi$ are polar and azimuthal angles respectively in conventional spherical polar coordinates(see Fig. \ref{fig1}). The vector potential has the following form,
\begin{equation}
    \vec{A} = A_0[\cos\left(\omega t\right)\vec{e_1} + \eta \sin\left(\omega t\right)\vec{e_2}],
\end{equation}
where, $A_0$ and $\omega$ represent the amplitude of the vector potential and frequency of the CPL, respectively. The parameter $\eta$ regulates the orientation of the polarization ($\eta=+1(-1)$ for left (right) CPL). All the three unit vectors $\vec{e_1}$, $\vec{e_2}$, and $\vec{n}$ must be orthogonal to each other. Thus, we choose unit vectors $\vec{e_1} = (\cos\theta\cos\phi,\cos\theta\sin\phi,\-\sin\theta)$ and $\vec{e_2} = (\sin\phi,-\cos\phi,0)$. We derive an effective stroboscopic Hamiltonian invoking the high-frequency Floquet formalism (see \cite{SM}), given by following form,
\begin{align}
    H^F(k) = &\sum_{j=x,y,z}\left[(\cos k_j - M-A^2/2) \tau_z\sigma_0+ \lambda \sin k_j\tau_x\sigma_j \right] \nn
    &+\tau_0\left(\vec{n}.\vec{\sigma}\right) +i\,\delta\,\tau_x\sigma_0,
\end{align}
where, the vector $\vec{n}$ is given by,
\begin{equation}
    \vec{n} = \frac{\eta\lambda^2A^2}{\omega}(\sin\theta\cos\phi,\sin\theta\sin\phi,\cos\theta).
\end{equation}

\begin{figure*}[t]
    \centering
\begin{tabular}{c c c c}
     \includegraphics[width=0.23\linewidth]{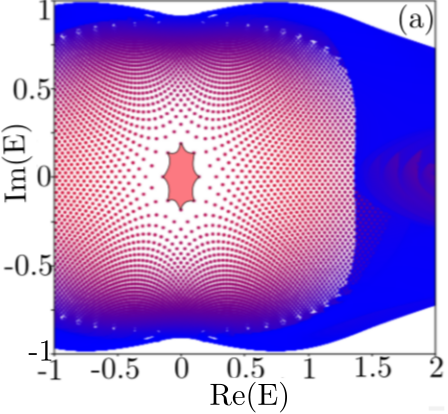}
    & \includegraphics[width=0.23\linewidth]{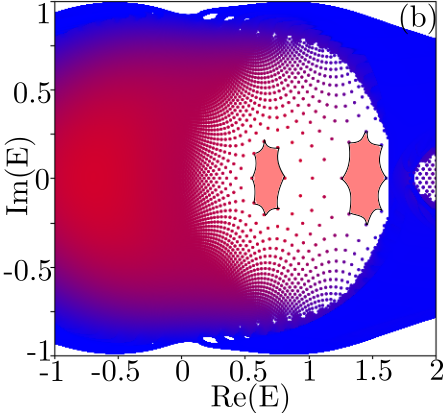}
     \hspace{+0.cm}&\includegraphics[width=0.23\linewidth]{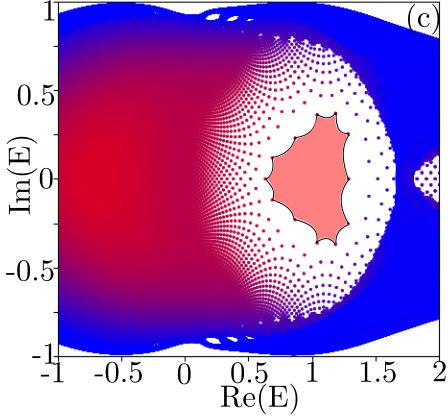}
     &\includegraphics[width=0.27\linewidth]{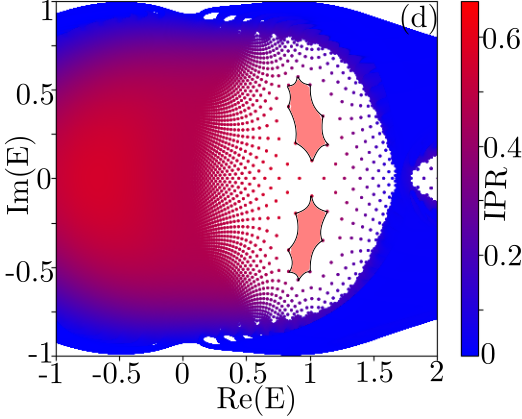} 
\end{tabular}  
    \caption{For one axis OBC along the $z$ axis, (a) depicts a signature of second order EP in the static phase for $M=3$, $\lambda = 1$, $\delta=1$, $B_0 = 0.2$ by an octagon shape. (b)-(d) represents the dynamics of the EP via Floquet driving the 3DNHTI in the presence of the magnetic field. We periodically vary $\frac{A_0^2}{\omega}\sin{\chi}$ and magnetic field as  $B_{0}\cos{\chi}$ for $\chi = 0.6, \, 0.65, \, 0.7$ in (b), (c), and (d) respectively. As $\chi$ varies, the two second-order EPs collide with each other and form a third-order EP which further splits into two second-order EPs. Other parameters are the same as (a) along with $M=2.5$ and $A_0^2/\omega=1/5$.}
    \label{fig3}
\end{figure*}

Here, the term coupled to the unit direction $\vec{n}$ can be defined as the  photo-induced magnetic field generated by the light-matter interaction as a result of the Floquet driving \cite{fqhe}.
This pseudo-magnetic field is analogous to artificial gauge fields realized in ultra-cold atoms but differs in the sense that, the pseudo-magnetic field components are the functions of parameters of irradiated light into the system which is generally anisotropic. The magnitude of the pseudo-magnetic field scales inversely to the frequency of the drive and its direction can be altered by changing the orientation of the polarization.

Thus, the stroboscopic phase represents a FETI resulting from the periodic driving of static 3DNHTI in the high-frequency limit.  As the time-reversal symmetry of the Hamiltonian is broken due to the pseudo-magnetic field, the model hosts a pair of chiral Weyl nodes along any desired axis with a suitable choice of $\theta$ and $\phi$ and gets connected over the imaginary axis due to NH term, thus giving rise to the point gap. Hence, for FETI, the system hosts  a point gap for $\lvert M+\frac{A^2}{2}-3\rvert\leq\delta$ [see Fig. \ref{fig2}(a)] and a line gap for $\lvert M+\frac{A^2}{2}-3\rvert\geq\delta$, see \cite{SM}. The Floquet Hamiltonian has $A^2/2$ onsite term in the diagonal, thereby modulating $M$, which is responsible for the band inversion. Therefore, the FETI phase can be realized in the system even if its static counterpart is in the trivial phase.

\textit{Surface States.\textemdash}
We define the critical angles of polarization as, $\theta_c=tan^{-1}(\sqrt{2})$ and $\phi_c = \pi/4$. This makes the pseudo-magnetic field isotropic along [111] direction. At these critical values of the angle, in the complex energy spectrum, the point gap is filled isotropically with the doubly degenerate surface states having a single Fermi point, shown in Fig. \ref{fig2}(b). We also consider another unconventional model where two Fermi points can appear, see \cite{SM}. Therefore, it exhibits a single-sheet structure.  The degenerate partners of surface states are localized to the opposite edges of the truncated lattice. The surface state has a dispersion relation $k_x + i k_y$, whereas, its degenerate partner has the dispersion relation $k_y + i k_x$. Thus, we infer that the single sheet establishes homotopy with 2D torus-shaped Brillouin zone (BZ), see \cite{SM} for more details. Additionally,  the surface states appear as a Dirac cone in the $k_x-k_y$ plane as shown in Figs. \ref{fig2}(c) and (d). However, the photo-induced homotopy does not remain futile from the infinitesimal perturbation from the critical angle. We also note that the Hamiltonian hosts a hugely defective infernal point at $k_x = k_y = 0$ in the complex eigenspectrum which leads to numerical instability in the static model, see \cite{SM}. However, for the dynamic case due to the pseudo-magnetic field, this infernal point converts into doubly degenerate non-defective points. 

\textit{Effect of external magnetic field\textemdash} We consider the system in the presence of an external magnetic field, by adding the term $\tau_z (\vec{B_{0}}\cdot \vec \sigma) $ to the Hamiltonian written in Eq.(\ref{1}) and Floquet Hamiltonian also changes to $H^{F}+ \tau_z (\vec{B_{0}} \cdot \vec{\sigma})$.  Thus, a constant magnetic field in the static phase gains dynamics in the stroboscopic limit. The resultant magnetic field modulates the imbalance between lande-g-factors between the $s$ and $p$ orbital. Thus we can photo-modulate the Land\'e-$g$-factor by parameterizing Floquet term as $\frac{\lambda^2 A_0^2}{\omega}\sin{\chi}$ and the magnetic field $B_{0}\cos \chi$ with
\begin{equation}
    \chi = \tan^{-1}\left(\frac{\lambda A_{0}^{2}}{\omega B_0}\right).
\end{equation}

The bulk Hamiltonian has a point gap however, the finite system along the $z$ axis, possesses a single second-order EP on the surface for the non-zero magnetic field accounting for a quantum anomaly on the surface, see Fig. \ref{fig3}(a). We note that if there is $n$th order EP, then one has to take $4n$  turns around it to come back to the original state in the square momentum grid \cite{denner2021exceptional}. The dynamics and the order of the EP residing on the surface can be photo-tuned by Floquet driving which can never be realized in its static counterpart with a constant unidirectional magnetic field. By adiabatically varying $\chi$, two second-order EPs come closer and collide with each other, thereby creating a third-order EP. This third-order EP can then be further transformed into two second-order by moving them away from each other (see Figs. \ref{fig3}(b)-(d)).

\begin{figure}[h]
    \centering
\begin{tabular}{c c c c}
      \includegraphics[width=0.25\linewidth]{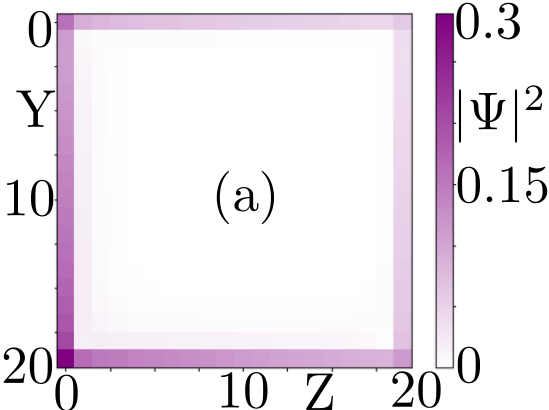}
     \includegraphics[width=0.25\linewidth]{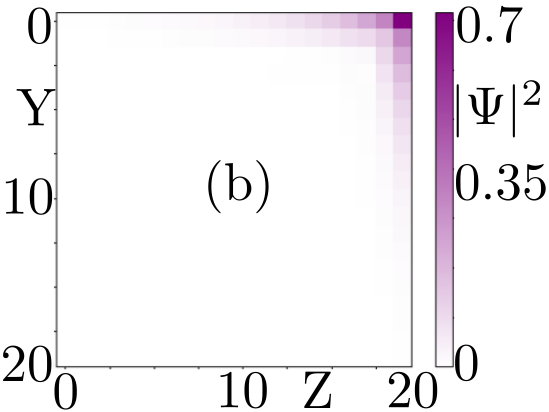}
      \includegraphics[width=0.25\linewidth]{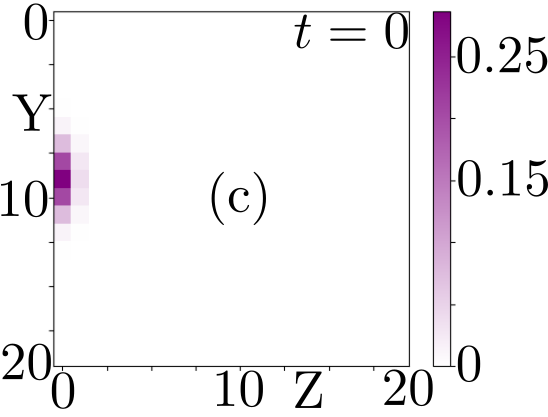}
     \includegraphics[width=0.25\linewidth]{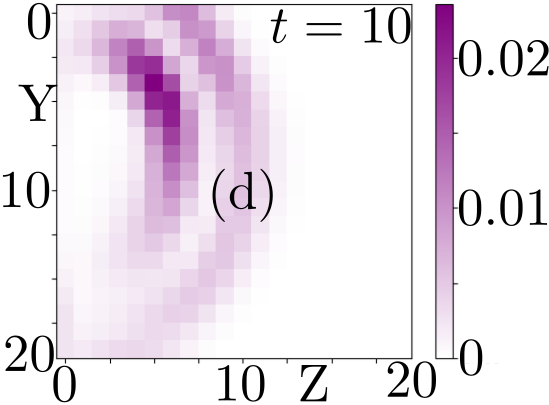}\\
      \includegraphics[width=0.25\linewidth]{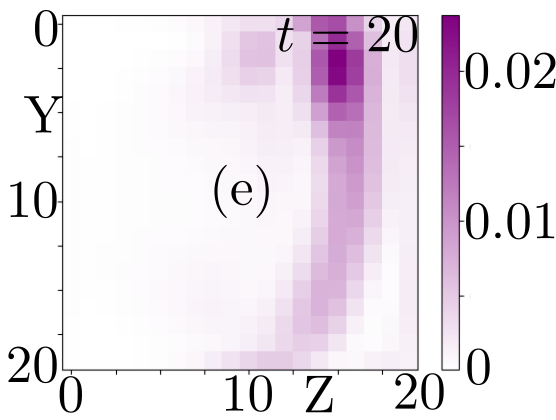}
     \includegraphics[width=0.25\linewidth]{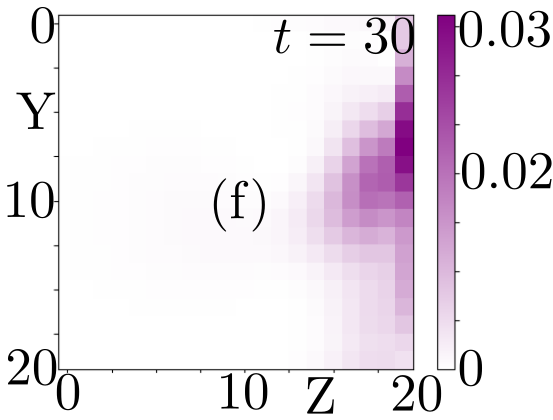}
     \includegraphics[width=0.25\linewidth]{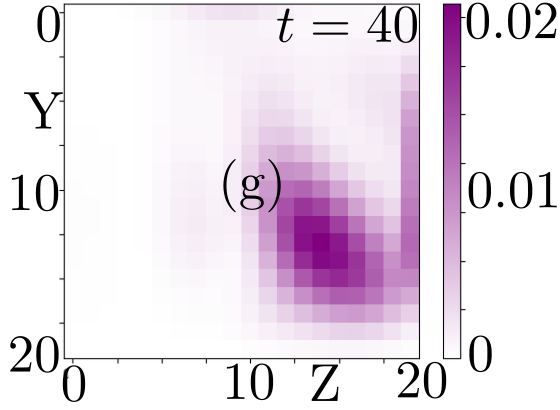}
     \includegraphics[width=0.25\linewidth]{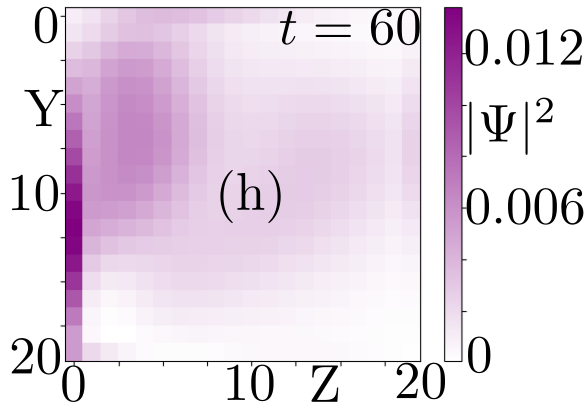}   
\end{tabular}   
    \caption{Probability amplitudes demonstrate the NHSEE 
    for (a) $B_0 = 0$ and (b) $B_0 = 0.2$. (c)-(f) demonstrates the dynamical evolution of the trial wavefunction from time $t=0$ to $t=60$. The constants $\alpha$ and $\beta$ are set to be 2 and 6 respectively. We choose the spinor $|\zeta_0\rangle$ as $[1,i,0,0]$, and the momentum $k_x$ is evaluated numerically considering those values which experience NHSSE. Other system parameters are the same as Fig. \ref{fig3}}
    \label{fig4}.
\end{figure}

We then apply open boundary condition (OBC) along $y$ and $z$, while retaining periodic BC along the $x$ axis. In the absence of the magnetic field, the surface states in the point gaps are localized in opposite corners. This remarkable phenomenon is the well-known NH surface skin effect (NHSSE) and can be related to the higher-order skin effect \cite{Zhong, Slanger}. However, as the magnetic field is turned on, the surface states get confined in one of the  corners (refer Figs. \ref{fig4}(a),(b)). To demonstrate this novel effect, we prepare a trial wavefunction localized at a finite $y-z$ sheet of dimension $L^2$. The trial wavefunction is given by:
\begin{eqnarray}
     |\psi_0\rangle = N_0 e^{-\frac{(y-y_0)^2}{\alpha^2}}e^{-\frac{(z-z_0)^2}{\beta^2}}
    e^{ik_x a_x}|\zeta_0\rangle
\end{eqnarray}
where, $N_0$ is the normalization factor of the wave function, $y_0$ and $z_0$ are constants that determine the localization of the wave function in the finite sheet, $\alpha$ and $\beta$ control its Gaussian width, and $|\zeta_0\rangle$ is a spinor. We perform the time evolution of the trial wavefunction as $e^{-iHt}|\Psi_0\rangle$ with respect to the Hamiltonian obtained by truncating $y$ and $z$ axes.
The skin modes are not completely localized at the corners and they have noticeable overlap with the bulk modes (Figs. \ref{fig4}(a)-(b)). So this effect allows the wavepacket which was initially localized  at one of the edges, to travel into the opposite edge by permeating into the bulk (Figs. \ref{fig4}(c)-(f)). However, there is a significant dynamical quenching of the time-evolved probability of the trial wavefunction at the edges where surface states are localized.

\begin{figure}[h]
    \centering
\begin{tabular}{c c}
      \includegraphics[width=0.48\linewidth]{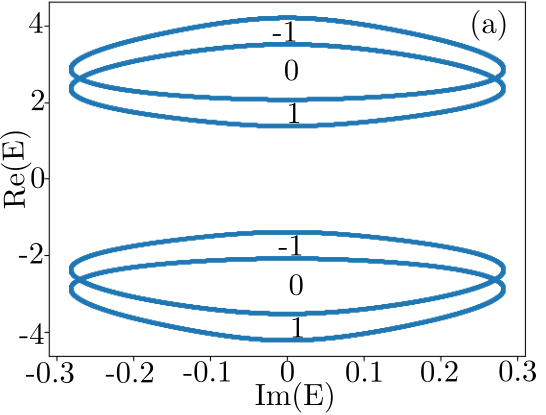}
      &\raisebox{-0.2cm}{\includegraphics[width=0.53\linewidth]{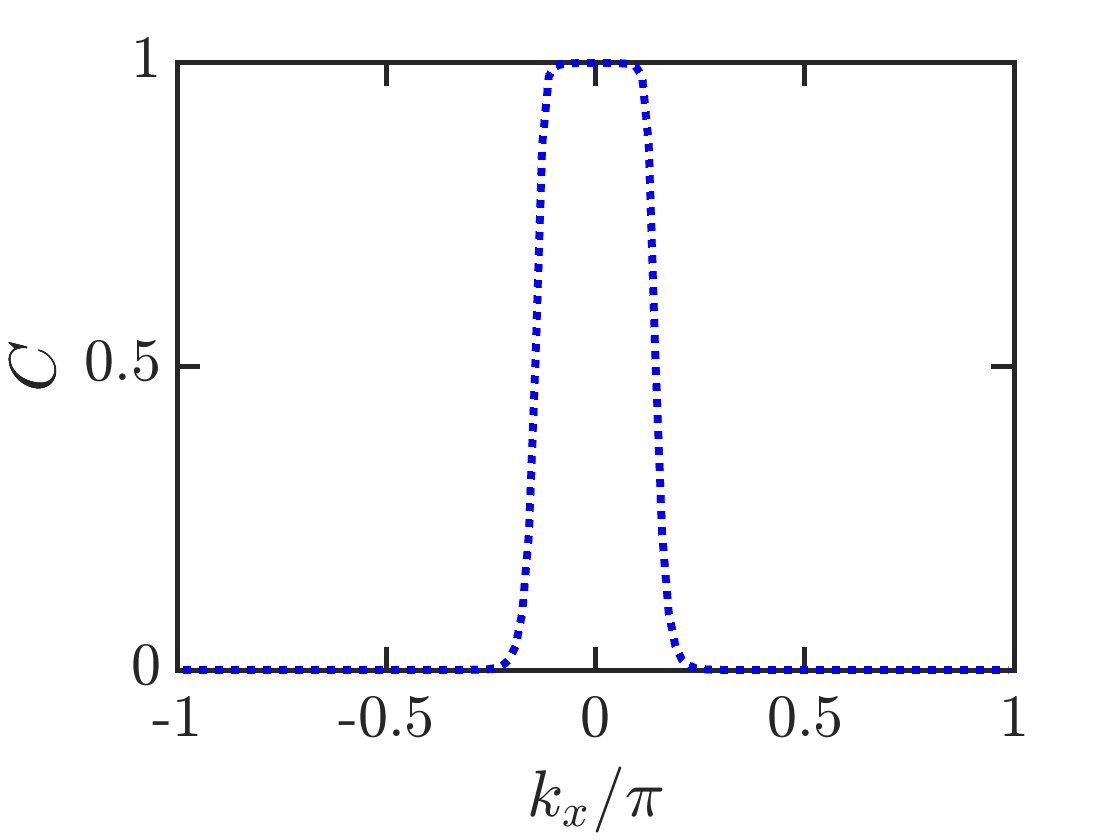}}\\
     \end{tabular}
     \caption{(a) The quantized spectral winding number in the complex eigenspectrum plane for $k_{x0} = -\pi$ and $k_{y0} = -\pi$. 
     (b) represents the Floquet biorthogonal Chen number depicting the quantized value of one in the region where there are surface states. System parameters are the same as in the previous figure with $B_0=0$.}      
    \label{fig5}  
\end{figure}

\textit{Topological Invariants.\textemdash}
We compute three topological invariants starting with the spectral winding number along a particular momentum axis say $k_z$ while fixing the other two momenta values at $k_{x0}$ and $k_{y0}$.   Thereafter, along the $k_z$ axis, the Hamiltonian becomes one-dimensional and can be defined as $H_{1D}^F(k_z)= H^F(k_{x0},k_{y0},k_z)$. We calculate the spectral winding number as,
\begin{eqnarray}
     \nu_{k_{x0},k_{y0}}(E_p) = \frac{1}{2\pi i}\int_{-\pi}^{\pi}dk_z \text{Tr}[Q^{1D}(k_z)],
\end{eqnarray}

where $Q^{1D}(k_z)=[H_{1D}^F(k_z)-E_p]^{-1}\partial_{k_z}[H_{1D}^F(k_z)-E_p]$ with $E_p$ as a reference energy inside the corresponding spectral region. Thus, for a suitable choice of complex reference energy in the eigenspectrum, the quantized value of the spectral winding number implies broken bulk-boundary correspondence (BBC) resulting an NH skin effect for those particular $k_{x0}$ and $k_{y0}$ along $k_z$ direction. As shown in Fig. \ref{fig5}(a), there are four bands with different values of $\nu$ along with the zero-valued region  that can be visualized as the outcome of adding two $\nu$'s in the overlapping complex eigenspectrum.

Furthermore, along with three spectral winding numbers in three directions  \cite{NHSEO}, system has three 1D winding numbers $W_{1D,l}$ defined as,
\begin{equation}
    W_{1D,l} = -i\int\frac{d^3k}{(2\pi)^3}Tr[Q_l(k)].
\end{equation}

where $Q_l=[H^F(k)-E]^{-1}\partial_{k_l}[H^F(k)-E]$ with $E$ being the reference energy in the point gap. For our system,   $W_{1D,l}=0$  for all values of $l$, which also necessitates the absence of NHSE \cite{Corres,NHSEO}. Thus, the uniqueness of the model is evident from the fact that the system as a whole does not exhibit the collapse of the BBC although the momentum slice in the Hamiltonian experience NHSE. There is also presence of another 3D topological invariant $W_{3D}$\cite{Sym,TopPh,Ghatak_2019,Toppds} defined in the bulk spectrum which is given by,

\begin{equation}
    W_{3D} = \frac{-1}{24\pi^2}\int d^3k\epsilon^{ijk}Tr[Q_i(k)Q_j(k)Q_k(k)].
\end{equation}

 The topological invariant defined in the bulk determines the fate of the surface states. The quantization of $W_{3D}$, however, does not require any symmetry for its stabilization.
Finally, we also employ a biorthogonal approach to compute the Floquet open-boundary Chen number for each $k_x$ value \cite{song}. This unique approach, however, re-establishes the BBC. The Floquet open-boundary Chern number is given by;

\begin{equation}
    C_{\alpha} = \frac{2\pi i}{l'_{y}l'_{z}} \text{Tr}'\left(\hat P_{\alpha}[[\hat r_{y},\hat P_{\alpha}],[\hat r_{z},\hat P_{\alpha}]]\right).
\end{equation}

Here $\hat r_y$ ($\hat r_z$) is the coordinate operator along $y$ ($z$) direction and defined as $\hat r_{y(z)_{mn}} = r_{y(z)}\delta_{mn}$ with $ r_x, r_y \le l $, $l\times l$ is the size of the system and $l'_{y}=l'_{z}=l-2l_0$ where $l_0$ is a boundary layer that has been removed from $l_{x/y}$, see \cite{SM}. The bulk band projection operator, $P_\beta=\sum_{n\in \beta}|nR\rangle \langle nL|$, where $\beta$ denotes all the unoccupied bands with $|nL\rangle$ ($|nR\rangle$) being the left (right) eigenstates of Floquet Hamiltonian. The Floquet biorthogonal Chern number gives the quantized value of one where there are surface states in the Hamiltonian [see Figs. \ref{fig2}(c) and \ref{fig4}(b)] and can be modulated by varying $\frac{A_0^2}{\omega}$. 

In conclusion, we have studied 3D NHTI in the presence of CPL. The system hosts a novel phase of quantum matter, namely, FETI in the stroboscopic limit with no static counterpart.  FETI has a point gap that is filled by either a single band at the surface or a 2D band with EPs when 3D NHTI is subjected to  an external magnetic field. Due to the NH fermion doubling theorem, an odd number of Fermi points or EPs are impossible in a 2D model. That's why the surface states in FETI are anomalous. The exceptional topology i.e. the number of EPs can also be modulated using CPL. We also established using different topological invariants, that the NH skin effect does not exist in the entire system, but it is present in momentum slices. Finally, the NHSSE exhibited by FETI has been explained by the dynamical quenching of the wave-function. Thus, a photo-induced modulation of the transport and quantum anomaly can also be realized in such a modeled NH system.\\
\textit{Acknowledgments\textemdash}  For financial support, S.B. thanks CSIR, India and
M.T. thanks Science and Engineering Research Board (India) 
grant SRG/2022/001408 and Young Faculty Incentive Fellowship from IIT Delhi. The authors would like to thank Ravi Gilani for computational resources.
\bibliography{sample.bib}
\newpage
\onecolumngrid
\begin{center}
{\bf \large{SUPPLEMENTARY MATERIAL: Floquet Exceptional Topological Insulator}} \\
\vspace*{0.2cm}
 Gaurab Kumar Dash, Subhajyoti Bid, Manisha Thakurathi \\
{\it Department of Physics, Indian Institute of Technology Delhi, Hauz Khas, New Delhi, India 110016}
\end{center}
\setcounter{equation}{0}
\setcounter{figure}{0}
\setcounter{table}{0}
\setcounter{page}{1}
\makeatletter
\renewcommand{\theequation}{S\arabic{equation}}
\renewcommand{\thefigure}{S\arabic{figure}}
\newcounter{SIfig}
\renewcommand{\theSIfig}{S\arabic{SIfig}}
\section{3D Non-Hermitian Topological Insulator}

\begin{figure}[h]
    \centering
\begin{tabular}{c c c c}
     \includegraphics[width=0.25\linewidth]{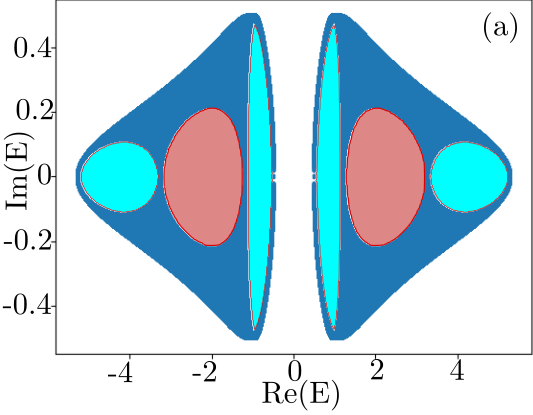}
     &\includegraphics[width=0.21\linewidth]{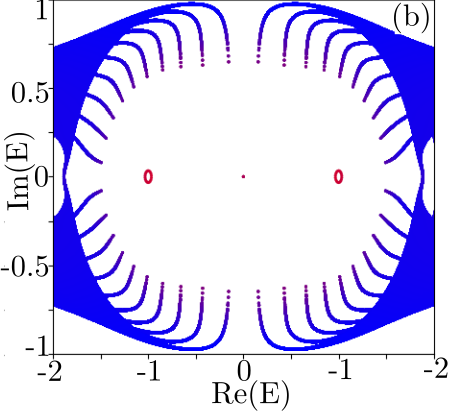}
     &\includegraphics[width=0.21\linewidth]{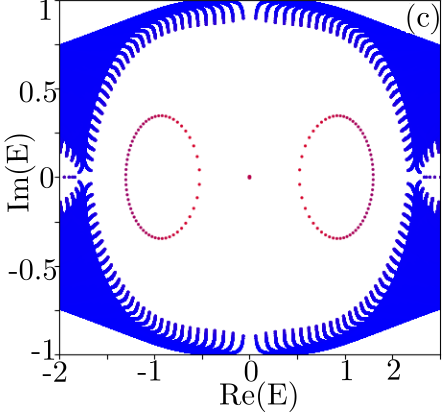}
     
     &\includegraphics[width=0.25\linewidth]{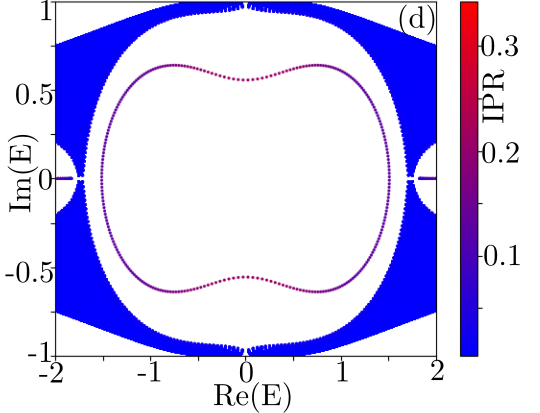}
    
 \end{tabular}         
    \caption{(a) shows a line gap for $M=2.3$, $\lambda = 1$, $\delta=0.5$, $B_0 = 0$. (b)-(d) shows the complex eigenspectrum of the Hamiltonian defined in Eq. S1 for lattice site N = 10, 30, and 70 respectively accounting for the numerical instability of the system.}
    \refstepcounter{SIfig}\label{supfig1}
\end{figure}

The Hermitian counterpart of the Hamiltonian $H_0(k)$ written in the main text is a four-band model of 3DTI hosting a Dirac node. For $1\leq|M|\leq3$, the system exhibits a trivial phase. The phase transition from trivial to topological phase occurs at $|M|=1$ and $|M|=3$. Due to the NH property, the Hamiltonian is accompanied by two kinds of gaps in the complex eigenspectrum\cite{denner2021exceptional,Dynweyl,RevModPhys.93.015005}. For $|M-3|\leq\delta$ and $|M-3|\geq\delta$, it showcases a central point gap and a central line gap respectively [refer Fig. \ref{supfig1}(a)].

The Hamiltonian is hugely defective in the surface for $k_x = k_y = 0$ and leads to numerical instability (refer Fig. \ref{supfig1}(b)-(e)). Thus, the model exhibits an infernal point in the thermodynamic limit at $k_x = k_y = 0$, which accounts for the states to be localized at one of the edges. An analytical derivation of the dispersion relation of the infernal point is presented in  \cite{denner2021exceptional} 

\section{Lattice Realization of 3DNHTI}
The Hamiltonian in the main text can be realized in a cubic lattice with an electron with spin up and down in $s$ and $p$ orbital respectively. Thereafter, the tight binding Hamiltonian is given by,

\begin{equation}
    H = \sum_{r,\gamma}C_{r,\gamma}^{\dagger}H(k)C_{r,\gamma},
\end{equation}

where, $r = {x,y,z}$ denotes the position of lattice and $\gamma = 0(1)$ notifies the $s(p)$ orbitals. We try to expand each term on the basis of the lattice mentioned in the main text. The constant part of the onsite term in the second quantization notation is written as,

\begin{eqnarray}
 {\sum_{r,\gamma}C_{r,\gamma}^{\dagger}(-M)\tau_{z}\sigma_{0}C_{r,\gamma} }
 &&= \sum_{r,\gamma}\begin{pmatrix}C_{r,s,\uparrow}^{\dagger} & C_{r,s,\downarrow}^{\dagger} & C_{r,p,\uparrow}^{\dagger} & C_{r,p,\downarrow}^{\dagger}
\end{pmatrix}\begin{pmatrix}
-M & 0 & 0 & 0 \\
0  & -M & 0 & 0 \\
0 & 0 & M & 0  \\
0 & 0 &  0 & M
\end{pmatrix}\begin{pmatrix}
C_{r,s,\uparrow} \\
C_{r,s,\downarrow} \\ 
C_{r,p,\uparrow} \\
C_{r,p,\downarrow}
\end{pmatrix} \nn
&&= \sum_{r} [-M C_{r,s,\uparrow}^{\dagger}C_{r,s,\uparrow} - MC_{r,s,\downarrow}^{\dagger}C_{r,s,\downarrow} -MC_{r,p,\uparrow}^{\dagger}C_{r,p,\uparrow} -MC_{r,p,\downarrow}^{\dagger}C_{r,,\downarrow}] \nonumber \\
&&= {-M\sum_{r,\gamma}(-1)^{\gamma}C_{r,\gamma}^{\dagger}\sigma_0C_{r,\gamma}}
\end{eqnarray}

where, $C_{r,\gamma}^{\dagger} = \begin{pmatrix}
C_{r,\gamma,\uparrow}^{\dagger} &    C_{r,p,\gamma, \uparrow}^{\dagger}
\end{pmatrix}$.

The $k$ dependent onsite terms are of the following form,
\begin{eqnarray}
 {\sum_{r,\gamma}C_{r,\gamma}^{\dagger}(\cos k_x)\tau_{z}\sigma_{0}C_{r,\gamma} }
 = &&\sum_{r}\begin{pmatrix}
C_{r,s,\uparrow}^{\dagger} & C_{r,s,\downarrow}^{\dagger} & C_{r,p,\uparrow}^{\dagger} & C_{r,p,\downarrow}^{\dagger}
\end{pmatrix}\begin{pmatrix}
\cos k_x & 0 & 0 & 0 \\
0  & \cos k_x & 0 & 0 \\
0 & 0 & -\cos k_x& 0  \\
0 & 0 &  0 & -\cos k_x
\end{pmatrix}\begin{pmatrix}
C_{r,s,\uparrow} \\
C_{r,s,\downarrow} \\ 
C_{r,p,\uparrow} \\
C_{r,p,\downarrow}
\end{pmatrix} \nn
&&= \sum_{r} [\cos k_xC_{r,s,\uparrow}^{\dagger}C_{r,s,\uparrow} - \cos k_xC_{r,s,\downarrow}^{\dagger}C_{r,s,\downarrow} -\cos k_xC_{r,p,\uparrow}^{\dagger}C_{r,p,\uparrow} -\cos k_xC_{r,p,\downarrow}^{\dagger}C_{r,p,\downarrow} ]\nn
&& = \frac{1}{2}\sum_{r}(C_{r+e_x,s,\uparrow}^{\dagger}C_{r,s,\uparrow} + C_{r+e_x,s,\downarrow}^{\dagger}C_{r,s,\downarrow} -C_{r+e_x,p,\uparrow}^{\dagger}C_{r,p,\uparrow} -C_{r+e_x,p,\downarrow}^{\dagger}C_{r,p,\downarrow})\nn
& &= {\frac{1}{2}\sum_{r,\gamma}(-1)^{\gamma}C_{r+e_x,\gamma}^{\dagger}\sigma_0C_{r,\gamma} + H.C}.
\end{eqnarray}

Hence, the collective term $\cos k_x + \cos k_y + \cos k_z$ can be evaluated as:
\begin{equation}
  \sum_{r,\gamma}C_{r,\gamma}^{\dagger}(\cos k_x+ \cos k_y + \cos k_z )\tau_{z}\sigma_{0}C_{r,\gamma} =   \frac{1}{2}\sum_{r,\gamma}\sum_{i=x,y,z}(-1)^{\gamma}C_{r+e_i,\gamma}^{\dagger}\sigma_0C_{r,\gamma} + H.C.
\end{equation}

Similarly, following the same steps of the calculation, the rest of the terms can be converted as:
\begin{equation}
  \sum_{r,\gamma}\sum_{i=x,y,z}C_{r,\gamma}^{\dagger}(\sin k_i)\tau_{x}\sigma_{i}C_{r,\gamma} =   \frac{\lambda}{2i}\sum_{r,\gamma}\sum_{i=x,y,z}(-1)^{\gamma}C_{r+e_i,\gamma + 1}^{\dagger}\sigma_iC_{r,\gamma} + H.C,
\end{equation}

\begin{equation}
  \sum_{r,\gamma}\sum_{i=x,y,z}C_{r,\gamma}^{\dagger}(B\tau_{z}\sigma_{i})C_{r,\gamma} =   B\sum_{r,\gamma}\sum_{i=x,y,z}(-1)^{\gamma}C_{r,\gamma }^{\dagger}\sigma_iC_{r,\gamma} + H.C,
\end{equation}
and
\begin{equation}
 \sum_{r,\gamma}\sum_{i=x,y,z}C_{r,\gamma}^{\dagger}i\delta\tau_{x}\sigma_{0}C_{r,\gamma} =   i\delta\sum_{r,\gamma}(-1)^{\gamma}C_{r,\gamma + 1}^{\dagger}\sigma_0C_{r,\gamma}.
\end{equation}

After collecting the terms the lattice Hamiltonian takes the following form,

\begin{eqnarray}
 H = && -M\sum_{r,\gamma}(-1)^{\gamma}C_{r,\gamma}^{\dagger}\sigma_0C_{r,\gamma} + [\frac{1}{2}\sum_{r,\gamma}(-1)^{\gamma}C_{r+e_x,\gamma}^{\dagger}\sigma_0C_{r,\gamma} + H.C] \nn
 &&+ \frac{\lambda}{2i}\sum_{r,\gamma}\sum_{i=x,y,z}(-1)^{\gamma}C_{r+e_i,\gamma + 1}^{\dagger}\sigma_iC_{r,\gamma} + H.C. \nn
&& + B\sum_{r,\gamma}\sum_{i=x,y,z}(-1)^{\gamma}C_{r,\gamma }^{\dagger}\sigma_iC_{r,\gamma} + H.C. + i\delta\sum_{r,\gamma}(-1)^{\gamma}C_{r,\gamma + 1}^{\dagger}\sigma_0C_{r,\gamma},
\end{eqnarray}

where, the NH part in the Hamiltonian can be realized as the electron-phonon interaction between $s$ and $p$ orbital.

\section{Formalism for Floquet Theory}
We define a non-unitary time evolution operator $U(t,t^{'})$ which evolves the system from time $t$ to $t{'}$ with periodicity $\tau = \frac{2\pi}{\omega}$, then the Floquet theorem states that,
\begin{equation}
    U(t+n\tau,t_0) = U(t,t_0)[U(t_0+\tau,t_0)]^{n}
\end{equation}

and we define the non-unitary Floquet Hamiltonian as:
\begin{equation}
    U(t_0+\tau,t_0) = exp(\frac{iH_{F}\tau}{\hbar})
\end{equation}

where, $H_F$ is the Floquet NH Hamiltonian. For the stroboscopic analysis, we can always set $t_0 = 0$, without the loss of generality, as the original time period of the periodically driven system dominates any time scale that may be acquired by the unitary operator. The eigenstates corresponding to the Floquet operators defined above for time $t_0 = 0$ are called the Floquet modes and are given by $\psi_{\alpha}(0)$. Thus, the Floquet operator can be rewritten as\cite{bukov2015universal},

\begin{equation}
    U(\tau,0) = \sum_{\alpha}e^{-i\epsilon_{\alpha}\tau/\hbar}\ket{\psi_{\alpha}(0)}\bra{\psi_{\alpha}(0)},
\end{equation}

where, $\epsilon_\alpha$ are the complex eigenvalues corresponding to the Floquet NH Hamiltonian\cite{wang2021non}. The Floquet modes also satisfy the periodic relation as, $\psi_{\alpha}(\tau)=\psi_{\alpha}(0)$. The time evolution of the Floquet modes is given by:
\begin{equation}
    \Psi_{\alpha}(t) = e^{-i\epsilon_{\alpha}t/\hbar}\psi_{\alpha}(0).
\end{equation}

Substituting the time-dependent ansatz into the time-dependent Schrodinger's equation we get,
\begin{equation}
    [H(t) - i\hbar\frac{\partial}{\partial t}]\Psi_{\alpha}(t) = \epsilon_{\alpha} \Psi_{\alpha}(t),
\end{equation}

where, $K(t) =H(t) - i\hbar\frac{\partial}{\partial t} $ can be termed as Floquet extended Hamiltonian.
\subsection{Floquet-space-time representation of periodically driven systems}
We can further expand the Floquet modes in the time-periodic Fourier series as:
\begin{equation}
    \Psi_{\alpha}(t) = \sum_{j=-\infty}^{\infty}\ket{\phi_{\alpha,j}}e^{ij\omega t}.
\end{equation}

Plugging this into the above equation yields:

\begin{equation}
    \sum_{j=-\infty}^{\infty}H_{j-j^{'}}\phi_{\alpha,j} + j\hbar\omega\phi_{\alpha,j} = \epsilon_{\alpha}\phi_{\alpha,j},
\end{equation}

where, 
\begin{equation}
    H_{j-j^{'}} = H_n = \frac{1}{\tau}\int_{-\tau/2}^{\tau/2}H(k,t)e^{in\omega t}.
\end{equation}

Thus, the above equation can be written in matrix formulation as:

\begin{equation}
\large{
    H_{slab} = \begin{bmatrix}
    H_0 & H_{-1} & 0 & \dots & 0 & 0 & 0\\
    H_{1} & H_0-\hbar\omega & H_{-1} & \dots & 0 & 0 & 0 \\
    0 & H_{1} & H_0-2\hbar\omega & \dots & 0 & 0 & 0 \\
    \vdots & \vdots & \vdots & \ddots & \vdots &\vdots &\vdots \\
    0 & 0 & 0 & \dots & \ddots & H_{-1} & 0 \\
    0 & 0 & 0 & \dots & H_{1} & \ddots & H_{-1} \\
    0 & 0 & 0 & \dots & 0 & H_{1} & H_0-j \hbar\omega\\
    \end{bmatrix}.
    }
\end{equation}

\begin{figure}
    \centering
    \includegraphics[width=0.4\linewidth]{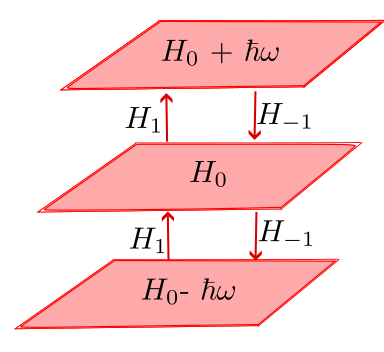}
    \caption{The Floquet variable onsite terms as the slabs and temporal hoppings are shown by red arrows}
    \refstepcounter{SIfig} \label{figs2}
\end{figure}

Thus, the harmonic indices $j$ and $j^{'}$ represent the fictitious temporal direction such that a $d$-dimensional Hamiltonian can always be visualized in the $d+1$ dimensional space-time representation(refer Fig. \ref{figs2}). The term $j\hbar\omega$ represents the variable onsite term (similar to a 1D chain under stark electric field) and $H_{j-j}^{'}$ represents the hopping between $j$ and $j^{'}$ temporal site. In addition to this, quasi-energy also satisfies the periodic relation of $\epsilon_{\alpha} = \epsilon_{\alpha} +j\hbar\omega $. Thus the space-time picture can also equally be mapped into the Wannier-Stark ladder\cite{avron1994periodic}.

considering the periodicity of the drive to be very high, we can write the Hamiltonian as :
\begin{equation}
    H = H_{0} + H_{1}e^{i\omega t} + H_{-1}e^{-i\omega t}.
\end{equation}
In the infinite frequency domain, the hopping along the temporal direction becomes completely ineffective breaking a $ d+1$-dimensional system (4-dimensional system in our case) to the isolated $d$-dimensional systems. However, in the low energy approximation, the perturbation theory yields unique results in the second order. If $\epsilon$ is the energy which is associated with the zeroth mode level($H_0$), then going from $n=0$ and $n=1$ and coming back is described by the term $H_{-1}\frac{1}{(\epsilon+\hbar\omega)-\epsilon}H_{1}^{\dagger}$ and going from $n=0$ to $n=-1$ and coming back is described by the term $H_{1}\frac{1}{(\epsilon-\hbar\omega)-\epsilon}H_{-1}^{\dagger}$. Thus, the whole process described above can be mathematically written as:
\begin{equation}
    H_{eff} = H_{F}^{0} + \sum_{n=1}^{\inf} \frac{1}{\omega}[H_{F}^{-n},H_{F}^{n}].
\end{equation}
\section{Floquet driving of 3DNHTI}
We use the above-developed perturbation theory in the conventional 3DNHTI and try to develop  FETI. We use the vector potential(mentioned in the main text) as:
\begin{equation}
    A = A_{0}[\cos\omega t \vec ~e_1 + \eta\sin\omega t \vec ~e_2],
\end{equation}

where, $A_0$ is the amplitude and $\omega$ is the frequency of the driven system. $\eta=\pm 1$ signifies the right circularly and left circularly polarized light respectively. We choose $e_1 = (\cos\theta\cos\phi,\cos\theta\sin\phi,\-sin\theta)$ and $e_2 = (\sin\phi,-\cos\phi,0)$. Thus, the vector potential is given by:
\begin{equation}
    A = A_{0}\left(\cos\theta\cos\phi\cos\omega t + \eta\sin\phi\sin\omega t,\cos\theta\sin\phi\cos\omega t - \eta\cos\phi\sin\omega t,-\sin\theta\cos\omega t\right),
\end{equation}

with minimal coupling the time-dependent Hamiltonian becomes,
\begin{eqnarray}
    &H(k) = \sum_{j}(\cos k_j - M-\frac{A_0^2}{4}(1+\eta^2))\tau_z\sigma_0 + \lambda\sum_{j}[\sin k_x - A_0(\cos\theta\cos\phi\cos\omega t + \eta\sin\phi\sin\omega t)\tau_x\sigma_x \nn
    & + \sin k_y -A_0(\cos\theta\sin\phi\cos\omega t - \eta\cos\phi\sin\omega t)\tau_x\sigma_y +\sin k_z -A_0(-\sin\theta\cos\omega t)\tau_x\sigma_z] + i\delta\tau_x\sigma_0.
\end{eqnarray}

Thus, the various Floquet terms can be recovered as:
\begin{equation}
    H_0 = \sum_{j}(\cos k_j - M - \frac{A_0^2}{4}(1+\eta^2))\tau_z\sigma_0 + \lambda\sum_{j}\sin k_j\tau_x\sigma_j + i\delta\tau_x\sigma_0,
\end{equation}

\begin{equation}
    H_{1} = -\frac{\lambda A_0}{2}([\cos\theta\cos\phi+i\eta\sin\phi]\tau_x\sigma_x + [\cos\theta\sin\phi-i\eta\cos\phi]\tau_x\sigma_y + \sin\theta\tau_x\sigma_z),
\end{equation}
and
\begin{equation}
    H_{-1} = -\frac{\lambda A_0}{2}([\cos\theta\cos\phi-i\eta\sin\phi]\tau_x\sigma_x + [\cos\theta\sin\phi+i\eta\cos\phi]\tau_x\sigma_y + \sin\theta\tau_x\sigma_z).
\end{equation}

Thus, in the Fourier space-time representation, each onsite term represents a 3DNHTI with onsite loss and gain term whereas the hopping between two connected 3DNHTI can be modulated by the angle of polarization. Thus the whole system can be modulated by $A_0,\theta,\phi$ to realize exotic quantum phases of matter.

We neglect the periodic fluctuating terms with coefficient $\frac{A_{0}^{2}}{\omega}$. Thus, by using equation 13 the effective Hamiltonian can be calculated as:
\begin{equation}
    H(k) = \sum_{j}(\cos k_j - M-\frac{A^2}{4}(1+\eta^2))\tau_z\sigma_0 + \lambda\sum_{j}\sin k_j\tau_x\sigma_j + \tau_0(n.\sigma) + i\delta\tau_x\sigma_0,
\end{equation}

where, the vector n is given by :
\begin{equation}
    n = \frac{\lambda^2A^2\eta}{\omega}(\sin\theta\cos\phi,\sin\theta\sin\phi,\cos\theta).
\end{equation}

However, in the presence of the extrinsic magnetic field, the Hamiltonian in the stroboscopic phase is given by:
\begin{equation}
\boxed{
    H(k) = \sum_{j}(\cos k_j - M-\frac{A^2}{4}(1+\eta^2))\tau_z\sigma_0 + \lambda\sum_{j}\sin k_j\tau_x\sigma_j + (n\tau_0  +  B\tau_z).\sigma + i\delta\tau_x\sigma_0}.
\end{equation}

\section{Lattice Realization of FETI}
The FETI can also be calculated in the cubic lattice as:

\begin{eqnarray}
& H = -(M + \frac{A_0^2}{2})\sum_{r,\gamma}(-1)^{\gamma}C_{r,\gamma}^{\dagger}\sigma_0C_{r,\gamma} + [\frac{1}{2}\sum_{r,\gamma}(-1)^{\gamma}C_{r+e_x,\gamma}^{\dagger}\sigma_0C_{r,\gamma} + H.C] \nn
& + [\frac{\lambda}{2i}\sum_{r,\gamma}\sum_{i=x,y,z}(-1)^{\gamma}C_{r+e_i,\gamma + 1}^{\dagger}\sigma_iC_{r,\gamma} + H.C] \nn 
& \sum_{r,\gamma}\sum_{i=x,y,z}(n_i + B(-1)^{\gamma})C_{r,\gamma}^{\dagger}\sigma_iC_{r,\gamma}+ i\delta\sum_{r,\gamma}C_{r,\gamma + 1}^{\dagger}\sigma_0C_{r,\gamma}.
\end{eqnarray}

Thus, irradiating light in a 3DNHTI generates Onsite excitation due to the coupling of light-matter interaction which is analogous to a photo-induced magnetic field. The uniqueness of such a fictitious magnetic field reveals its true nature from the fact that it is generally anisotropic(except at the critical angle defined in the main text). The amplitude of such a field is dependent on the frequency and handedness of the light used.

\section{Photo-induced Homotopy in FETI}
We truncate $x$-axis while retaining PBC along $y$ and $z$ axes. For simplicity, we use the angle of polarization as $\theta=0$ and $\phi = \phi '$ where $0\leq\phi '\leq 2\pi$. Then the truncated hamiltonian can be written as:
\begin{eqnarray}
    & H = \frac{1}{2}[\sum_{x}C_{x+1}^{\dagger}(\tau_z\sigma_0 + i\lambda\tau_x\sigma_x)C_{x} + \sum_{x}C_{x-1}^{\dagger}(\tau_z\sigma_0 - i\lambda\tau_x\sigma_x)C_{x} ]\nn
    & + C_{x}^{\dagger}[(2-M-\frac{A_{0}^{2}}{2})]\tau_z\sigma_0 + \frac{\lambda^2 A_{0}^{2}}{\omega}\tau_0\sigma_z + i\delta\tau_x\sigma_0]C_{x}.
\end{eqnarray}

We use the trial wavefunction as:
\begin{equation}
    \boxed{\ket{\Psi_{\Bar{1}00}} = \sum_{x}\alpha^{x}\ket{x}\ket{\zeta_0}}
\end{equation}

We write the Harper's equation ($\Gamma_0 = \tau_z\sigma_0$) as\cite{Dynweyl},
\begin{equation}
    \boxed{\Gamma_0\left(\frac{1 - \lambda\tau_y\sigma_x}{2}\alpha^{-1} + \frac{1 + \lambda\tau_y\sigma_x}{2}\alpha + [(2-M-\frac{A_{0}^{2}}{2}) + \frac{\lambda^2 A_{0}^{2}}{\omega}\tau_z\sigma_z - \delta\tau_y\sigma_0]\right)\ket{\zeta_0} = 0}
\end{equation}

The terms $\tau_z\sigma_z$ and $\tau_y\sigma_0$ commutes with $\tau_y\sigma_x$.Thus the eigenstates of $\tau_y\sigma_x$ are given by:
\begin{equation}
    \ket{\psi_1} = \frac{(-i,0,0,1)^{T}}{\sqrt{2}}
\end{equation}
\begin{equation}
    \ket{\psi_2} = \frac{(0,-i,1,0)^{T}}{\sqrt{2}}
\end{equation}
\begin{equation}
    \ket{\psi_{-1}} = \frac{(i,0,0,1)^{T}}{\sqrt{2}}
\end{equation}
\begin{equation}
    \ket{\psi_{-2}} = \frac{(0,i,1,0)^{T}}{\sqrt{2}}
\end{equation}

Thus we write the spinor $\ket{\zeta_0}$ as:
\begin{equation}
    \ket{\zeta_0} = p_1\ket{\psi_1} + p_2\ket{\psi_2}
\end{equation}

We set $p_1 = \cos{\theta'}$ and $p_2 = \sin{\theta'}e^{i\phi '}$. Thus, after normalization, the Harper equation reduces to:
\begin{equation}
    \left(\alpha - M - \frac{A_0^2}{2} + \frac{\lambda^2 A_{0}^{2}}{\omega}\right)\cos{\theta '} - \delta\sin{\theta '}e^{i\phi '} = 0
\end{equation}
\begin{equation}
     \left(\alpha - M - \frac{A_0^2}{2} + \frac{\lambda^2 A_{0}^{2}}{\omega}\right)\sin{\theta '}e^{i\phi '} - \delta\cos{\theta '} = 0
\end{equation}

Thus the value of the constants are given by
\begin{equation}
    \alpha = \sqrt{(\delta^2 + (\frac{\lambda^2 A_0^2}{\omega})^2)} + (M + \frac{A_0^2}{2}) -2
\end{equation}
\begin{equation}
    \theta ' = \tan^{-1}(\frac{\alpha-M - \frac{A_0^2}{2} + 2 +\frac{\lambda^2 A_{0}^{2}}{\omega} }{\delta})
\end{equation}
\begin{equation}
    \phi' = 0
\end{equation}

Following the similar steps, we can write the trial wavefunction for (100) surface as :
\begin{equation}
    \boxed{\ket{\Psi_{\Bar{1}00}} = \sum_{x}\alpha^{x}\ket{x}\ket{\zeta'_0}}
\end{equation}

Then , after normalisation, the surface states are given by:
\begin{equation}
    \ket{\Psi_{\Bar{1}00}} = \sum_{x}\alpha^{L-x}\ket{x}[\cos{\theta'}\ket{\psi_{-1}} - \sin{\theta'}\ket{\psi_{-2}}]
\end{equation}
\begin{equation}
    \ket{\Psi_{100}} = \sum_{x}\alpha^{x}\ket{x}[\cos{\theta'}\ket{\psi_{-1}} + \sin{\theta'}\ket{\psi_{-2}}]
\end{equation}

For the system to establish homotopy with the torus shaped BZ and exhibits a single sheet in the surface, the top and buttom surfaces must couple to each other demanding the surface state in the complex eigenspectrum to be a superposition of $\ket{\Psi_{100}}$ and $\ket{\Psi_{\Bar{1}00}}$.Thus we assume the surface state to have the following form.
\begin{equation}
    \ket{\psi} = \frac{(\ket{\Psi_{100}})\pm(\ket{\Psi_{\Bar{1}00}})}{\sqrt{2}}.
\end{equation}

We then consider the perturbative correction of $k_y$ and $k_z$ respectively. Since $\bra{\psi_1}\tau_x\sigma_y\ket{\psi_1} = -\bra{\psi_2}\tau_x\sigma_y\ket{\psi_2} = \bra{\psi_{-1}}\tau_x\sigma_y\ket{\psi_{-1}} = \bra{\psi_{-2}}\tau_x\sigma_y\ket{\psi_{-2}} = 1$.Hence $\bra{\Psi_{\Bar{1}00}}\tau_x\sigma_y\ket{\Psi_{\Bar{1}00}} = - \bra{\Psi_{100}}\tau_x\sigma_y\ket{\Psi_{100}} = \cos{2\theta'}$, thus $k_y$ terms results in the energy splitting of $\pm\cos{2\theta'}k_{y}$ for small value of $k_y$ which would cage  the surface states to localise at one surface. Following the similar calculations for the $k_z$ term, we have, $\bra{\psi_1}\tau_x\sigma_y\ket{\psi_2} = -\bra{\psi_2}\tau_x\sigma_y\ket{\psi_1} = -\bra{\psi_{-1}}\tau_x\sigma_y\ket{\psi_{-2}} = \bra{\psi_{-2}}\tau_x\sigma_y\ket{\psi_{-1}} = i$. But $k_z$ dependent terms are not diagonal unlike the $k_y$ dependent terms. So, $\ket{\Psi_{100}}$ and $\ket{\Psi_{\Bar{1}00}}$ are not the good basis of the perturbed Hamiltonian when $k_z$ dependent perturbations are included in the system. To Tackle this difficulty, we re-solve the Harper's equation again without the NH terms.
\begin{equation}
     \boxed{\Gamma_0\left(\frac{1 - \lambda\tau_y\sigma_x}{2}\alpha^{-1} + \frac{1 + \lambda\tau_y\sigma_x}{2}\alpha + [(2-M-\frac{A_{0}^{2}}{2}) + \frac{\lambda^2 A_{0}^{2}}{\omega}\tau_z\sigma_z - \delta\tau_y\sigma_0 + i\sin{k_z}\tau_y\sigma_z]\right)\ket{\zeta_0} = 0}
\end{equation}
 As $[\tau_y\sigma_z,\tau_x\sigma_y]=0$,the $k_z$ dependent term inter-twins and mixes the Hilbert space of the unperturbed Hamiltonian since $\tau_y\sigma_z\ket{\psi_1} = \ket{\psi_{-2}}$, $\tau_y\sigma_z\ket{\psi_2} = \ket{\psi_{-1}}$, $\tau_y\sigma_z\ket{\psi_{-2}} = \ket{\psi_1}$, and $\tau_y\sigma_z\ket{\psi_{-1}} = -\ket{\psi_2}$.Thus for ($\Bar{1}00$) surface we redefine the spinor as 
 \begin{equation}
     \ket{\zeta} = \cos{\theta_1}\ket{\psi_1} + \sin{\theta_1}e^{i\phi_1}\ket{\psi_{-2}}
 \end{equation}

 Re-solving the Harper's equation we get
 \begin{equation}
     \alpha' = \frac{\beta_1 + \beta_2}{\beta_3}
 \end{equation}
 \begin{equation}
     \beta_1 = -1 + (\frac{\lambda^2 A_{0}^{2}}{\omega})^2 - (\sin{k_z})^2 - (M + \frac{A_0^2}{2})
 \end{equation}
 \begin{equation}
     \beta_2 = - \sqrt{(1-(\frac{\lambda^2 A_{0}^{2}}{\omega})^2 + (\sin{k_z})^2) + 4((\frac{\lambda^2 A_{0}^{2}}{\omega})^2 - (M + \frac{A_0^2}{2})^2) + (M + \frac{A_0^2}{2})^2}
 \end{equation}
 \begin{equation}
     \beta_3 = 2(M + \frac{A_0^2}{2} - \frac{\lambda^2 A_{0}^{2}}{\omega})
 \end{equation}
 \begin{equation}
     \theta_1 = -\tan^{-1}(\frac{\sin{k_z}}{\alpha' + M + \frac{A_0^2}{2} + \frac{\lambda^2 A_{0}^{2}}{\omega}})
 \end{equation}
 \begin{equation}
     \phi_1 = -\frac{\pi}{2}
 \end{equation}

 similarly for (100) surface, we can write
 \begin{equation}
     \ket{\zeta_1} = \cos{\theta_1}\ket{\psi_1} - i\sin{\theta_1}\ket{\psi_2}
 \end{equation}
 \begin{equation}
     \ket{\zeta_2} = \cos{\theta_1}\ket{\psi_{-1}} + i\sin{\theta_1}\ket{\psi_{-2}}
 \end{equation}
 Now including the NH electron-phonon interaction in the Hilbert space of $\ket{\zeta_{1,2}}$: $\bra{\zeta_1}i\tau_x\sigma_0\ket{\zeta_1} = \bra{\zeta_2}i\tau_x\sigma_0\ket{\zeta_2} = i\sin{2\theta_1}$. Thus the NH interaction splits the energy level equally. For small $k_z$, $\theta_1$ is directly proportional to $k_z$ and energy splitting in the NH complex energy is directly proportional to $ik_z$. Thus, the surface states have zero energy for finite $\delta$ since the NH term does not alter the surface states for $k_z = 0$ as $\theta_1 = 0$ when $k_z = 0$ thereby forming a single sheet in the complex eigenspectrum.
\section{Phase Space of FETI }
We neglect $\frac{A^2}{\omega}$ terms in the Hamiltonian to invoke the hidden sub-lattice symmetry of it by rotating the basis as, $\tau_x\Rightarrow\tau_y\Rightarrow\tau_z\Rightarrow\tau_x$. Then the Hamiltonian can be converted into the off-block diagonal form\cite{denner2021exceptional},

\begin{eqnarray}
& H(k,M,\lambda,\delta,A) = \sum_{j}(\cos k_j - M-\frac{A^2}{4}(1+\eta^2))\tau_x\sigma_0 + \lambda\sum_{j}\sin k_j\tau_y\sigma_j  + i\delta\tau_y\sigma_0 \nn
&\Rightarrow H(k,M,\lambda,\delta,A) = \begin{pmatrix}
0 & h_{UR}^{\dagger} \\
h_{LL} & 0 
\end{pmatrix},
\end{eqnarray}

where, $h_{UR}^{\dagger}$ and $h_{LL}$ denotes the upper right and lower left matrices respectively and given by,

\begin{equation}
    h_{UR}^{\dagger} = \begin{pmatrix}
f(k,M,A)-i\lambda\sin k_z + \delta & -i\lambda\sin k_x - \lambda\sin k_y \\
-i\lambda\sin k_x + \lambda\sin k_y & f(k,M,A)+i\lambda\sin k_z + \delta 
\end{pmatrix}    
\end{equation}
and
\begin{equation}
    h_{LL} = \begin{pmatrix}
f(k,M,A)+i\lambda\sin k_z - \delta & i\lambda\sin k_x + \lambda\sin k_y \\
i\lambda\sin k_x - \lambda\sin k_y & f(k,M,A)-i\lambda\sin k_z - \delta 
\end{pmatrix}    .
\end{equation}

Here, $f(k,M,A) = \sum_{j}(\cos{k_j} - M - \frac{A^2}{4}(1+\eta^2)$ we can then define $Q_i = H^{-1}\partial_{k_j}H= \begin{pmatrix}
h_{LL}^{-1}\partial_{k_i}h_{LL} & 0 \\
0 & (h_{UR}^{\dagger})^{-1}\partial_{k_i}h_{LL}
\end{pmatrix}$.

The topological invariant $W_{3D}$ can then be expressed as,

\begin{equation}
    W_{3D} = W_{3D}^{LL} - W_{3D}^{UR}.
\end{equation}
We can further link both of the off diagonal matrices as considering,
\begin{equation}
    h_{LL}(k,M,A,\lambda,\delta) = h_0(k,M + \frac{A^2}{4}(1+\eta^2)+\delta,\lambda),
\end{equation}

\begin{equation}
    h_{UR}^{\dagger}(k,M,A,\lambda,\delta) = h_0(k,M + \frac{A^2}{4}(1+\eta^2)-\delta,\lambda)
\end{equation}

and the Hamiltonian $h_0$ is written in the form,
\begin{equation}
    h_0 = d^{\mu}\sigma_{\mu}
\end{equation}

where, $d^{\mu} = (\cos k_j-M^{'},i\lambda\sin k_x, i\lambda\sin k_y, i\lambda\sin k_z)$ and $\sigma_{\mu} = (\sigma_0,\sigma_1,\sigma_2,\sigma_3)$. Thus it represents a two-band model of 3DTI with NH spin-orbit-coupling. For vanishing point gap the system demands that $h_0(k,M^{'},\lambda)=0$, implies, $\cos k_j = M^{'}$. This yields, $M^{'} = {\pm 1,\pm3}$ which decides the phase space of the block matrices.
Thus, the winding number can be expressed as,
\begin{equation}
    W_{3D} = W_{3D}^{0}(k,M + \frac{A^2}{4}(1+\eta^2)+\delta,\lambda) - W_{3D}^{0}(k,M + \frac{A^2}{4}(1+\eta^2)-\delta,\lambda),
\end{equation}

 and, $M^{'} = \pm 3$ implies $\mid M + \frac{A^2}{4}(1+\eta^2)\mid = 3.$ 
 Thus, $\mid M + \frac{A^2}{4}(1+\eta^2)-3\mid \leq \delta$ corresponds to a central point-gap whereas $\mid M + \frac{A^2}{4}(1+\eta^2)-3\mid \geq \delta$ corresponds to a central line-gap. Let us consider the model to be in static topological phase boundary for which it demands the value of $M=3$. Then by shining light on such a quantum phases of matter in the topological phase boundary, a phase transition from point gap to the line gap can be achieved. $ \frac{A^2}{4}(1+\eta^2) \leq \delta$ demands a central point gap and $ \frac{A^2}{4}(1+\eta^2) \geq \delta$ demands a central line gap.
 
\begin{figure}[h]
    \centering
\begin{tabular}{c c c c}
     \includegraphics[width=0.24\linewidth]{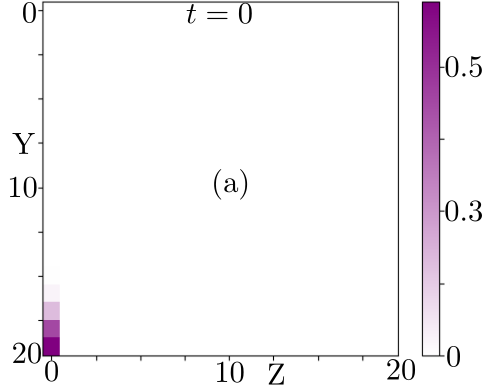}&\includegraphics[width=0.24\linewidth]{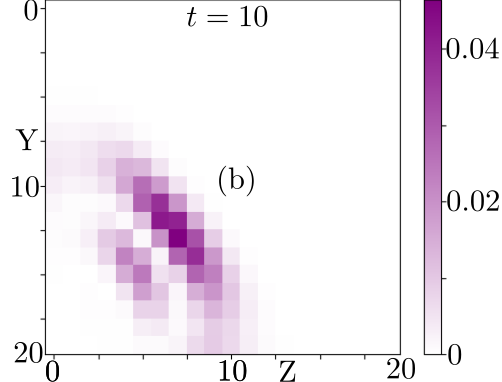}&\includegraphics[width=0.24\linewidth]{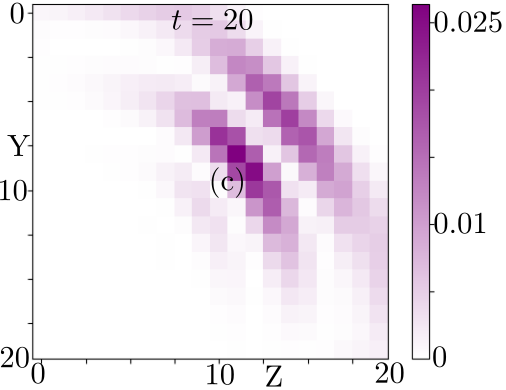}&\includegraphics[width=0.24\linewidth]{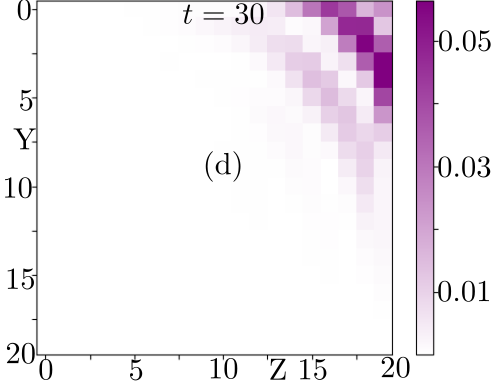} \\
     \includegraphics[width=0.24\linewidth]{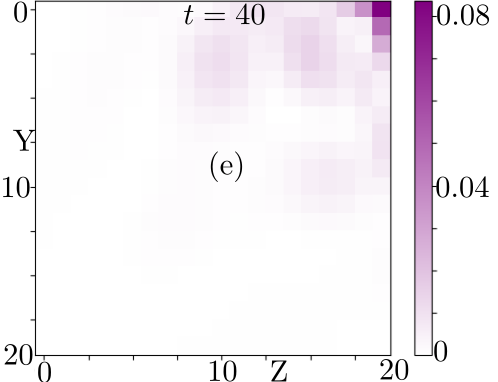}&\includegraphics[width=0.24\linewidth]{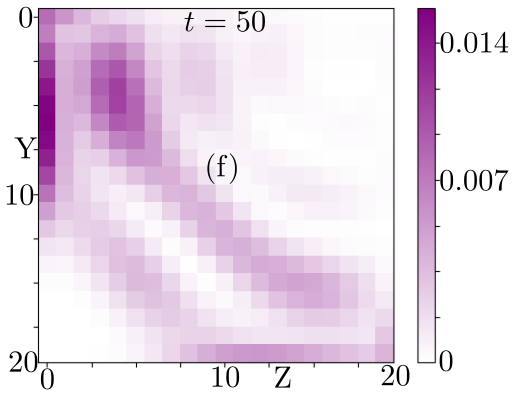}&\includegraphics[width=0.24\linewidth]{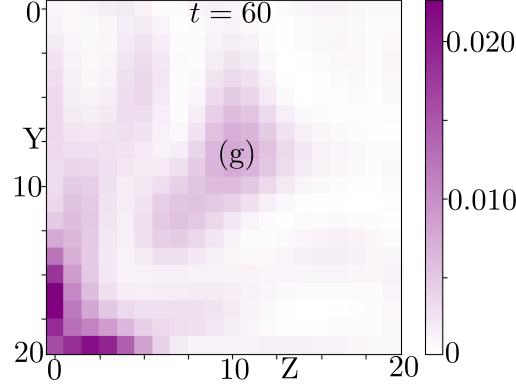}&\includegraphics[width=0.24\linewidth]{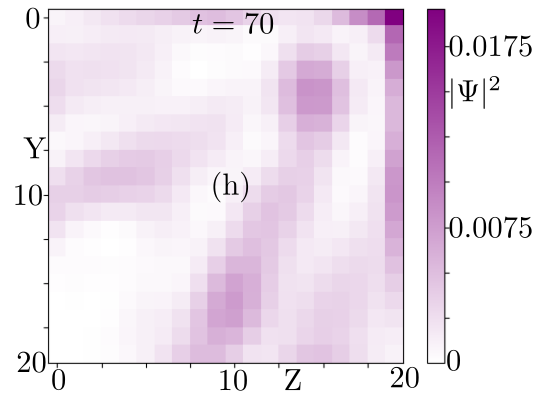} \\
     
     \centering

 \end{tabular}         
    \caption{(a)-(h) corresponds to the dynamic evolution of the probabilities of the trial wavefunction defined in the main text for $M=3$,$\delta=1$,$\lambda = 1$,$B_0 = 0.2$,from time $t=0$ to $t=70$ in the interval of 10. $\ket{\zeta_{0}}=[1,0,i,0]$ and the value of $\alpha$ and $\beta$ remains same as in the main text.}
     \refstepcounter{SIfig}\label{figs3}
\end{figure}

\section{Slab Calculation and wave dynamics}
We prepare a trial wave function, defined in the main text, and evolve it with respect to the slab Hamiltonian in which two of the axes are truncated. In the tight-binding representation, it is given by,

\begin{equation}
\boxed{
    H = \sum_{i,j}H_0C_{i,j}^{\dagger}C_{i,j} + \left(t_y C_{i,j}^{\dagger}C_{i+1,j} + t_zC_{i,j}^{\dagger}C_{i,j+1} +H.C\right)}
\end{equation}

where, $H_0$ is given by,
\begin{equation}
    H_0 = \sum_{j}\left(\cos k_j - M-\frac{A^2}{2}\right)\tau_z\sigma_0 + \lambda\sin k_y\tau_x\sigma_y +\tau_0\left(n.\sigma\right) + \tau_z\left(B.\sigma\right)+i\delta\tau_x\sigma_0.
\end{equation}
The hoppings takes the following form,
\begin{equation}
    t_y = \frac{1}{2}\left(\tau_z\sigma_0 -i\lambda\tau_x\sigma_x\right),
\end{equation}
\begin{equation}
    t_{y}^{\dagger} = \frac{1}{2}\left(\tau_z\sigma_0 +i\lambda\tau_x\sigma_x\right),
\end{equation}
\begin{equation}
    t_z = \frac{1}{2}\left(\tau_z\sigma_0 -i\lambda\tau_x\sigma_z\right),
\end{equation}
and
\begin{equation}
    t_{z}^{\dagger} = \frac{1}{2}\left(\tau_z\sigma_0 +i\lambda\tau_x\sigma_z\right).
\end{equation}

We have mentioned wave-dynamics evolution for critical angles in the main text. In this section,  we demonstrate the wave-dynamics evolution of the wave function for the static case in the presence of an external magnetic field. 

The wave-packet initially localized in one of the corners of the 2D sheet permeates into the bulk to travel to the opposite corner and returns to the same corner when the system evolved with respect to time. This evolution of an arbitrary Gaussian wave-packet account for the mimic phenomena of NHSSE exhibited by the model in the static phase with an external magnetic field where part of the wave-function probabilities are always quenched dynamically along the corners (see Fig.\ref{figs3}(a)-(h)). Thus, although the time evolution makes the wave-function to get absorbed into the bulk, it immediately escapes the bulk by penetrating to the other corner since the duration for which the it remains in the bulk is negligible compared to the duration for which it remains in the corners.

\section{Unconventional 3DNHTI}
We define an unconventional 3DNHTI as a cubic lattice of 3DNHTI (mentioned in the main text) with intrinsic spin orbit coupling (SOC) preserving the spin alignment controlled by the parameter $\Delta$. The rest of the terms has usual meaning as discussed for the previous model. The tight binding Hamiltonian is given by,
\begin{equation}
\boxed{
    H(k) = \sum_{j}(\cos k_j - M)\tau_z\sigma_0 + \lambda\sum_{j}\sin k_j\tau_x\sigma_j 
     + \Delta\sum_{j}\sin k_j\tau_y\sigma_0 + i\delta\tau_x\sigma_j}.
\end{equation}

\begin{figure}[h]
    \centering
\begin{tabular}{c c c}
     \includegraphics[width=0.28\linewidth]{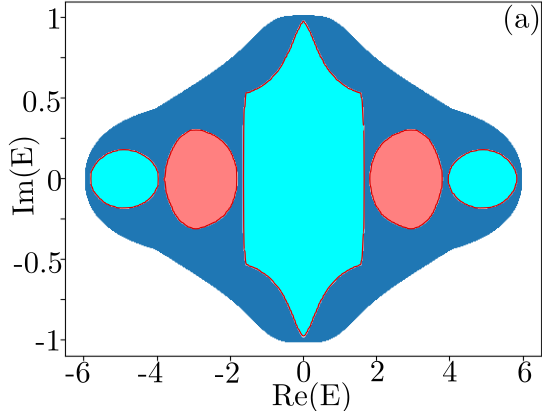}&\includegraphics[width=0.28\linewidth]{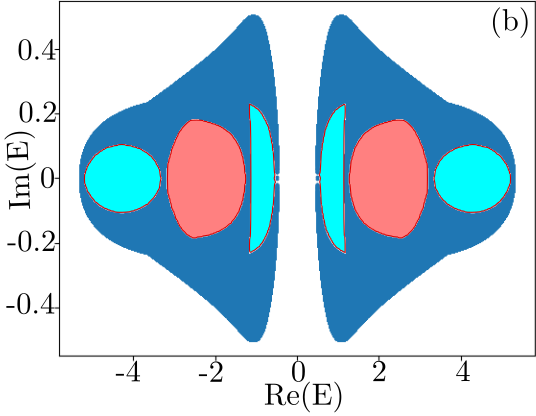}&\includegraphics[width=0.28\linewidth]{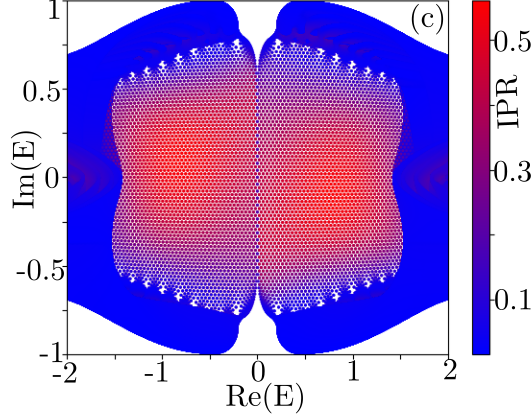}\\
     
     \centering
 \end{tabular}         
    \caption{(a) and (b) depicts the central point gap and line gap for $M=3$,$\lambda=1$,$\delta=1$,$\Delta=1$ and $M=2.3$,$\lambda=1$,$\delta=0.5$,$\Delta=1$ respectively.(c) shows a single sheet structure in the complex eigenspectrum for $M=2.5$,$\lambda=1$,$\delta=1$,$\Delta=1$,$A_0 = 1$,$\omega=5$ when the model is truncated along z-direction.}
     \refstepcounter{SIfig}\label{figs4}
\end{figure}

The model exhibits similar complex eigenspectrum as that of 3DNHTI. For $|M-3|\leq\delta$ and $|M-3|\geq\delta$, it showcases a central point gap and a central line gap respectively (refer Fig. \ref{figs4}(a)-(b)).It is, however, a band strained version of 3DNHTI near the BZ which accounts for the flatness of the bands at BZ boundary. However, the phase diagram of this unconventional 3DNHTI remains same as the 3DNHTI discussed above.

\section{Lattice Realization of Unconventional 3DNHTI}
We express its tight binding Hamiltonian in the cubic lattice as,

\begin{eqnarray}
& H = -M\sum_{r,\gamma}(-1)^{\gamma}C_{r,\gamma}^{\dagger}\sigma_0C_{r,\gamma} + [\frac{1}{2}\sum_{r,\gamma}(-1)^{\gamma}C_{r+e_x,\gamma}^{\dagger}\sigma_0C_{r,\gamma} + H.C] \nn
& + [\frac{\lambda}{2i}\sum_{r,\gamma}\sum_{i=x,y,z}(-1)^{\gamma}C_{r+e_i,\gamma + 1}^{\dagger}\sigma_iC_{r,\gamma} + H.C] \nn 
&  + [\frac{\Delta}{2}\sum_{r,\gamma}C_{r+e_x,\gamma +1}^{\dagger}\sigma_0C_{r,\gamma} + H.C] + + i\delta\sum_{r,\gamma}(-1)^{\gamma}C_{r,\gamma + 1}^{\dagger}\sigma_0C_{r,\gamma}.
\end{eqnarray}

\section{Floquet driving of unconventional 3DNHTI}
With the vector potential defined in equation S21, minimal coupling yields the time-dependent Hamiltonian,

\begin{eqnarray}
    &H(k) = \sum_{j}(\cos k_j - M-\frac{A_0^2}{4}(1+\eta^2))\tau_z\sigma_0 + \lambda[\sin k_x - A_0(\cos\theta\cos\phi\cos\omega t + \eta\sin\phi\sin\omega t)\tau_x\sigma_x \nn
    & + \sin k_y -A_0(\cos\theta\sin\phi\cos\omega t - \eta\cos\phi\sin\omega t)\tau_x\sigma_y +\sin k_z -A_0(-\sin\theta\cos\omega t)\tau_x\sigma_z] \nn
    & \Delta((\sin k_x - A_0(\cos\theta\cos\phi\cos\omega t + \eta\sin\phi\sin\omega t)) + (\sin k_y -A_0(\cos\theta\sin\phi\cos\omega t - \eta\cos\phi\sin\omega t))\nn
    & + (\sin k_z -A_0(-\sin\theta\cos\omega t)))\tau_y\sigma_0 + i\delta\tau_x\sigma_0.
\end{eqnarray}

The various Floquet modes are extracted as,

\begin{equation}
    H_0 = \sum_{j}(\cos k_j - M - \frac{A_0^2}{4}(1+\eta^2))\tau_z\sigma_0 + \lambda\sum_{j}\sin k_j\tau_x\sigma_j + \Delta\sum_{j}\sin k_j\tau_y\sigma_0  + i\delta\tau_x\sigma_0,
\end{equation}

\begin{eqnarray}
   & H_{1} = -\frac{\lambda A_0}{2}([\cos\theta\cos\phi + i\eta\sin\phi]\tau_x\sigma_x + [\cos\theta\sin\phi-i\eta\cos\phi]\tau_x\sigma_y \nn
   & + \sin\theta\tau_x\sigma_z) - -\frac{\Delta A_0}{2}([\cos\theta\cos\phi+i\eta\sin\phi] \nn
   & + [\cos\theta\sin\phi-i\eta\cos\phi]+\sin\theta)\tau_y\sigma_0,
\end{eqnarray}
and
\begin{eqnarray}
   & H_{-1} = -\frac{\lambda A_0}{2}([\cos\theta\cos\phi-i\eta\sin\phi]\tau_x\sigma_x + [\cos\theta\sin\phi+i\eta\cos\phi]\tau_x\sigma_y \nn
   & + \sin\theta\tau_x\sigma_z) - -\frac{\Delta A_0}{2}([\cos\theta\cos\phi-i\eta\sin\phi] \nn
   & + [\cos\theta\sin\phi+i\eta\cos\phi]+\sin\theta)\tau_y\sigma_0.
\end{eqnarray}

Thus, expanding the time dependent Hamiltonian in the stroboscopic phase yields:

\begin{equation}
    \boxed{H(k) = \sum_{j}(\cos k_j - M-\frac{A^2}{2})\tau_z\sigma_0 + \lambda\sum_{j}\sin k_j\tau_x\sigma_j  + \Delta\sum_{j}\sin k_j\tau_y\sigma_0 -\tau_0(n.\sigma)-\tau_z(n'.\sigma) + i\delta\tau_x\sigma_0},
\end{equation}
where, $n$ is the photodressed vector obtained above and $n'$ vector is given by,
\begin{equation}
    \boxed{n' = (\cos\theta-\sin\theta\sin\phi,\cos\theta-\sin\theta\cos\phi,\sin\theta (\cos\phi - \sin\phi))}.
\end{equation}

\begin{figure}[h]
    \centering
\begin{tabular}{c c c }
     \includegraphics[width=0.27\linewidth]{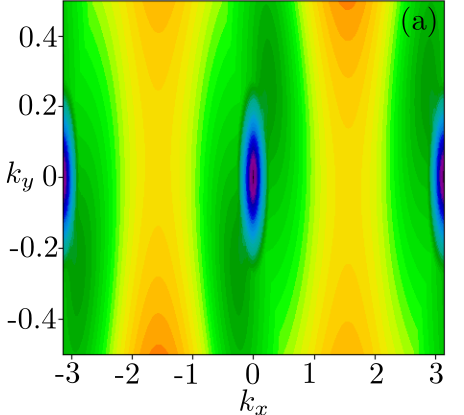}
     \hspace{0.2cm}
     &\includegraphics[width=0.26\linewidth]{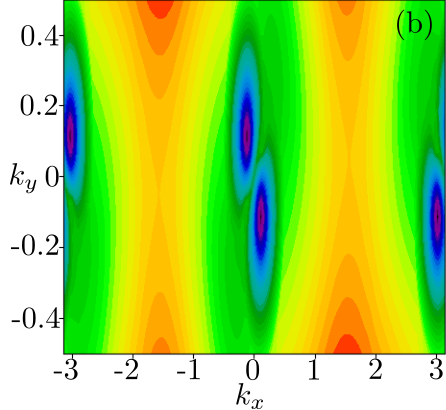}
     \hspace{0.2cm}
     
    &\includegraphics[width=0.33\linewidth]{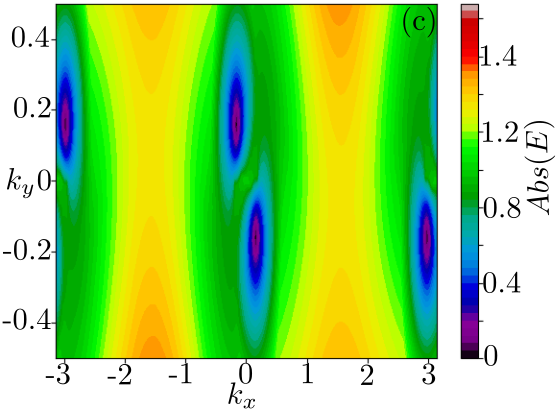}\\
     \centering 
 \end{tabular}         
    \caption{(a) denotes an $\text{Abs}(E)$ contour plot for $M=2.5$,$\lambda=1$,$\delta=1$,$\Delta=1$,$A_0 = 1$,$\omega=5$,$\theta=tan^{-1}\sqrt{2}$ and $\phi=\frac{\pi}{4}$.(b) and (c) show the sliding of two of the sheets $M=2.5$,$\lambda=1$,$\delta=1$,$A_0 = 1$,$\omega=5$,$\theta=\frac{\pi}{2}$ and $\phi=\frac{\pi}{4}$ by changing $\Delta$ as -0.6 and -0.9 respectively.}
     \refstepcounter{SIfig}\label{figs5}
\end{figure}

Thus, the Floquet driving generates two kinds of photo-dressed vectors that behave like the magnetic field. For the critical angle, the vectors become isotropic and exhibit a single sheet hosting two Fermi point(unlike the single sheet of FETI hosting a single Fermi point) and establish homotopy with the BZ (see Fig. \ref{figs4}(c) and \ref{figs5}(a)), but for the choice of the suitable angle of polarization, the single sheet evolves into a double sheet by sliding over each other hosting a non-degenerate edge state at each of the surfaces (see Fig. \ref{figs5}(b)-(c)). Therefore, the angle of polarization helps one to switch between both cases of the band spectrum on the surface. Hence $\Delta$ modulates the photo-dressed Land\'e-g-factor, in the sense, deciding the alignment of s and p orbital along the photo-dressed magnetic field even when the system does not experience any external magnetic field. The interaction which gives rise to additional spin-orbit coupling in the static phase and transforms a conventional 3DNHTI into an unconventional 3DNHTI also gives rise to the photo-dressed lande-g-factor for the dynamic case.

\section{Lattice realization of Floquet Unconventional 3DNHTI}
The FETI can be realized in the cubic lattice as,
\begin{eqnarray}
& H = -(M+\frac{A_0^2}{2})\sum_{r,\gamma}(-1)^{\gamma}C_{r,\gamma}^{\dagger}\sigma_0C_{r,\gamma} + [\frac{1}{2}\sum_{r,\gamma}(-1)^{\gamma}C_{r+e_x,\gamma}^{\dagger}\sigma_0C_{r,\gamma} + H.C] \nn
& + [\frac{\lambda}{2i}\sum_{r,\gamma}\sum_{i=x,y,z}(-1)^{\gamma}C_{r+e_i,\gamma + 1}^{\dagger}\sigma_iC_{r,\gamma} + H.C] \nn 
&  + [\frac{\Delta}{2}\sum_{r,\gamma}C_{r+e_x,\gamma +1}^{\dagger}\sigma_0C_{r,\gamma} + H.C] 
 + \sum_{r,\gamma}n_iC_{r,\gamma}^{\dagger}\sigma_iC_{r,\gamma} \nn
&+ \sum_{r,\gamma}(-1)^{\gamma}n_i^{'}C_{r,\gamma}^{\dagger}\sigma_iC_{r,\gamma} + i\delta\sum_{r,\gamma}C_{r,\gamma + 1}^{\dagger}\sigma_0C_{r,\gamma}.
\end{eqnarray}

\section{Slab Hamiltonian and Wave Dynamics Evolution}
As mentioned in Eq. 26, we determine the slab Hamiltonian and calculate the onsite and hopping part of the Hamiltonian. These are given by,

\begin{equation}
    H_0 = \sum_{j}\left(\cos k_j - M-\frac{A^2}{2}\right)\tau_z\sigma_0 + \lambda\sin k_y\tau_x\sigma_y +\tau_0\left(n.\sigma\right) +\tau_z\left(n^{'}.\sigma\right) + i\delta\tau_x\sigma_0,
\end{equation}
\begin{equation}
    t_y = \frac{1}{2}\left(\tau_z\sigma_0 -i\lambda\tau_x\sigma_y-i\Delta\tau_y\sigma_0\right),
\end{equation}
\begin{equation}
    t_{y}^{\dagger} = \frac{1}{2}\left(\tau_z\sigma_0 +i\lambda\tau_x\sigma_y + i\Delta\tau_y\sigma_0\right),
\end{equation}
\begin{equation}
    t_z = \frac{1}{2}\left(\tau_z\sigma_0 -i\lambda\tau_x\sigma_z - i\Delta\tau_y\sigma_0\right),
\end{equation}
and
\begin{equation}
    t_{z}^{\dagger} = \frac{1}{2}\left(\tau_z\sigma_0 +i\lambda\tau_x\sigma_z + i\Delta\tau_y\sigma_0\right)
\end{equation}

\begin{figure}[h]
    \centering
\begin{tabular}{c c c c}
     \includegraphics[width=0.24\linewidth]{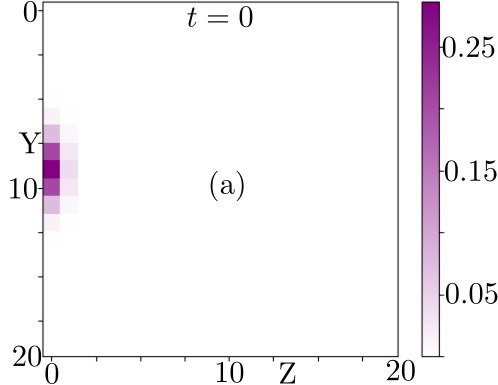}&\includegraphics[width=0.24\linewidth]{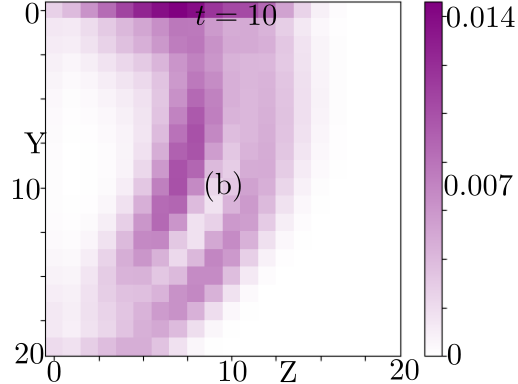}&\includegraphics[width=0.24\linewidth]{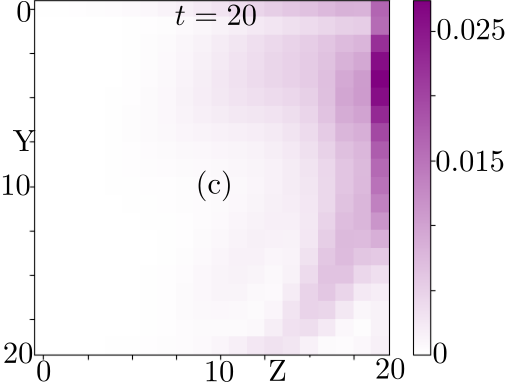}&\includegraphics[width=0.24\linewidth]{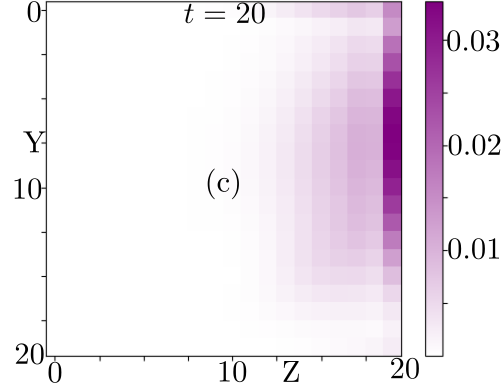} \\
     \includegraphics[width=0.24\linewidth]{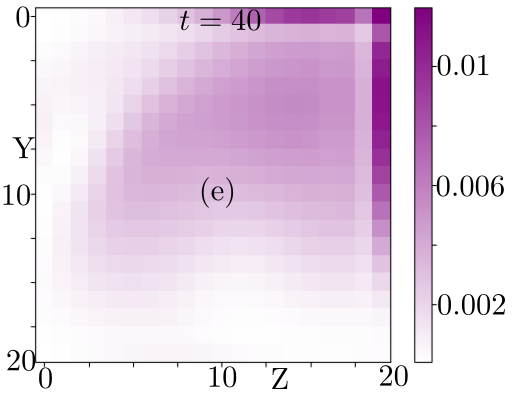}&\includegraphics[width=0.24\linewidth]{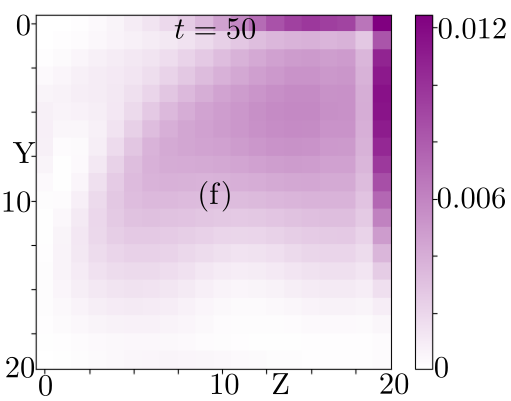}&\includegraphics[width=0.24\linewidth]{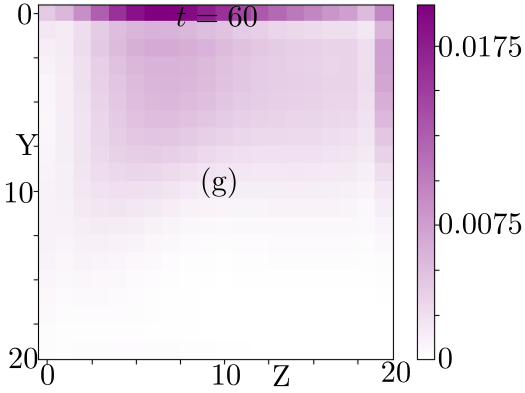}&\includegraphics[width=0.24\linewidth]{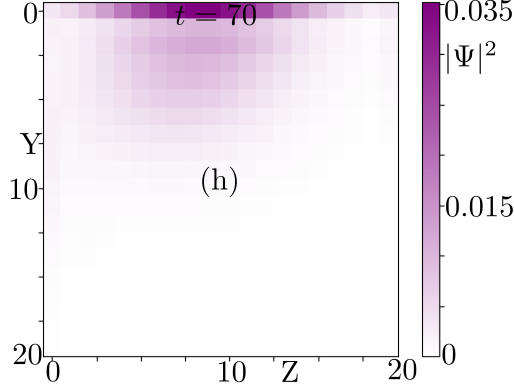} \\
     
     \centering

 \end{tabular}         
    \caption{(a)-(h) corresponds to the dynamic evolution of the probabilities of the trial wavefunction initially localized at the middle of one of the edges for $M=2.5$,$\delta=1$,$\lambda = 1$,$\Delta=1 = 0.2$,$A_0 = 1$,$\omega = 5$,$\theta=tan^{-1}\sqrt{2}$ and $\phi=\frac{\pi}{4}$ from time $t=0$ to $t=70$ in the interval of 10.$\ket{\zeta_{0}}=[1,i,i,0]$ and the value of $\alpha$ and $\beta$ remains same as in the main text.}
     \refstepcounter{SIfig}\label{figs6}
\end{figure}

\begin{figure}[h]
    \centering
\begin{tabular}{c c c c}
     \includegraphics[width=0.24\linewidth]{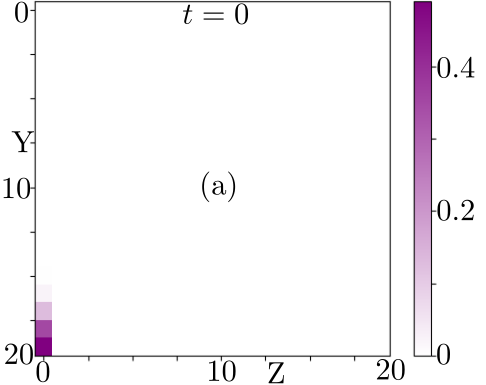}&\includegraphics[width=0.24\linewidth]{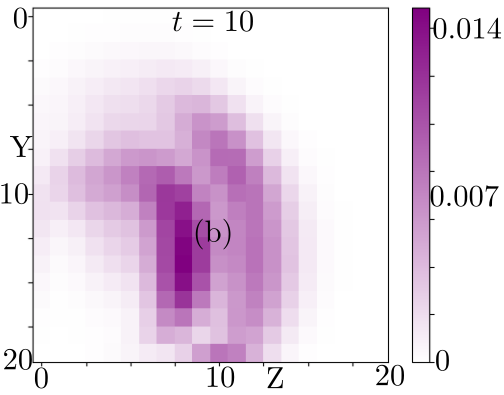}&\includegraphics[width=0.24\linewidth]{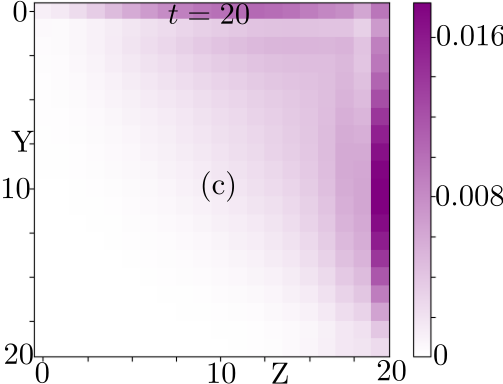}&\includegraphics[width=0.24\linewidth]{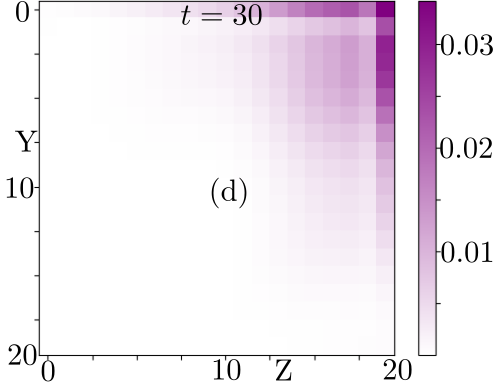} \\
     \includegraphics[width=0.24\linewidth]{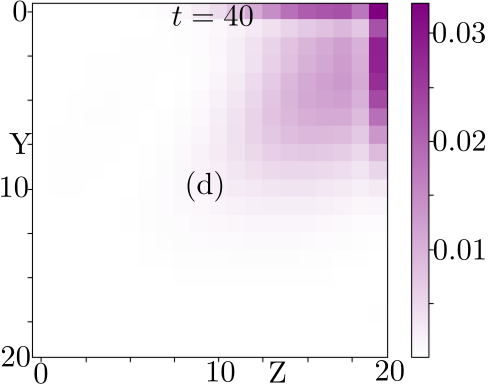}&\includegraphics[width=0.24\linewidth]{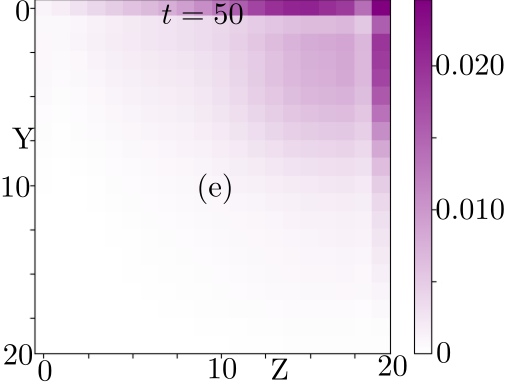}&\includegraphics[width=0.24\linewidth]{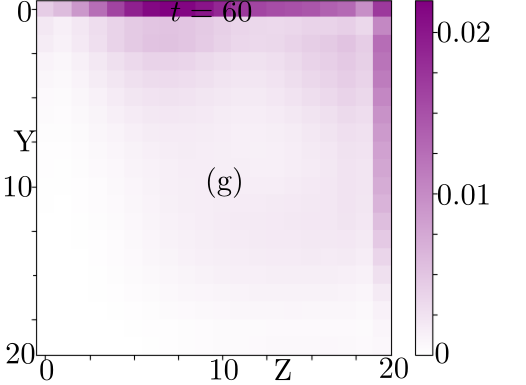}&\includegraphics[width=0.24\linewidth]{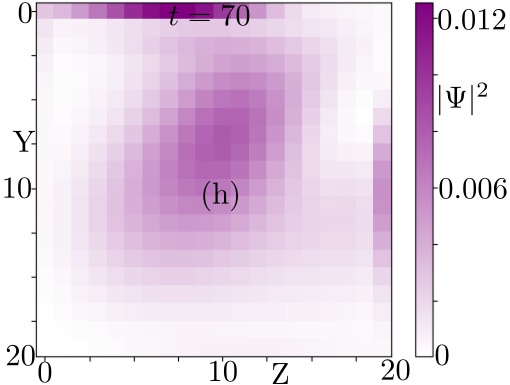} \\
     
     \centering

 \end{tabular}         
    \caption{(a)-(h) corresponds to the dynamic evolution of the probabilities of the trial wavefunction initially localized at one of the corners for $M=2.5$,$\delta=1$,$\lambda = 1$,$\Delta=1 = 0.2$,$A_0 = 1$,$\omega = 5$,$\theta=tan^{-1}\sqrt{2}$ and $\phi=\frac{\pi}{4}$ from time $t=0$ to $t=70$ in the interval of 10.$\ket{\zeta_{0}}=[1,i,i,0]$ and the value of $\alpha$ and $\beta$ remains same as in the main text.}
     \refstepcounter{SIfig}\label{figs7}
\end{figure}
respectively. For case (1), the surface states are localized in one of the edges and it has a significant amount of overlap integral with the bulk (refer Fig. \ref{figs6}(a)-(h)). So, the wavefunction initially localized in the middle of the Z-axis also has a significant amount of non-vanishing amplitude in the bulk which makes the localized wavefunction penetrate into the bulk and allow a part of it to travel along the edges. The wavepacket stays on the edge for the longer time where the surface states are caged.

For the second case, the wavefunction initially localized in one of the corners travels to the other corners and gets localized  with a quasi-static phase in the dynamic evolution (see Fig. \ref{figs7}(a)-(h)). The localization of the wavefunction increases at the corner where the majority of them are trapped. So, the wave dynamic evolution of the localized Gaussian wavefunction is guided by the NHSSE entrailed within the Hamiltonian which can be modulated by varying the angle of polarization of the CPL.

\end{document}